\documentclass[Afour,sageh,times]{sagej}
\usepackage{moreverb,url}
\usepackage[colorlinks,bookmarksopen,bookmarksnumbered,citecolor=red,urlcolor=red]{hyperref}

\usepackage{tabularx}
\usepackage{paralist}
\usepackage{color, colortbl}
\usepackage{algorithm}
\usepackage{algpseudocode}
\usepackage{algorithmicx}
\usepackage{multirow}
\usepackage{lineno}

\usepackage{subcaption}
\usepackage{caption}

\definecolor{Sepia}{rgb}{0.44, 0.26, 0.08}
\definecolor{Gray}{gray}{0.9}
\definecolor{Gray0}{gray}{0.95}
\definecolor{Gray2}{gray}{0.85}
\definecolor{Gray3}{gray}{0.8}

\newcommand{\ipoint}[1]{\textbf{\textit{#1}}}
\newcommand{\idef}[1]{{\color{Sepia}\textbf{\textit{#1}}}}
\newcommand{\bipa}{\begin{inparaenum}[(\itshape i\upshape)]}
\newcommand{\eipa}{\end{inparaenum}}
\newcommand{\bipasub}{\begin{inparaenum}[(\itshape a\upshape)]}
\newcommand{\eipasub}{\end{inparaenum}}
\newcommand{\MyPara}[1]{\textbf{#1}.}
\DeclareMathOperator{\bigO}{\mathcal{O}}
\newcommand{\igrad}{\ensuremath{\nabla}} 
\newcommand{\ilap}{\rotatebox[origin=c]{180}{$\nabla$}} 
\newcommand{\CodeIn}[1]{{\ttfamily\texttt{#1}}}
\newcommand{\revisedb}[1]{\textcolor{black}{#1}}

\newcommand\BibTeX{{\rmfamily B\kern-.05em \textsc{i\kern-.025em b}\kern-.08em T\kern-.1667em\lower.7ex\hbox{E}\kern-.125emX}}

\def\volumeyear{2025}
\begin{document}


\title{A Portable Multi-GPU Solver for Collisional Plasmas with Coulombic Interactions}

\author{James Almgren-Bell\affilnum{1}, Nader Al Awar\affilnum{1}, Dilip S Geethakrishnan\affilnum{1}, Milos Gligoric\affilnum{1}, and George Biros\affilnum{1}}

\affiliation{\affilnum{1}The University of Texas at Austin; Austin, 78712 TX, USA}

\corrauth{James Almgren-Bell}

\email{jalmgrenbell@utexas.edu}

\begin{abstract}
  We study parallel particle-in-cell (PIC) methods for low-temperature plasmas (LTPs), which discretize kinetic formulations that capture the time evolution of the probability density function of particles as a function of position and velocity. We use a kinetic description for electrons and a fluid approximation for heavy species. In this paper, we focus on GPU acceleration of algorithms for velocity-space interactions and in particular, collisions of electrons with neutrals, ions, and electrons. Our work has two thrusts. The first is algorithmic exploration and analysis. The second is examining the viability of rapid-prototyping implementations using Python-based HPC tools, in particular PyKokkos. We discuss several common PIC kernels and present performance results on NVIDIA Volta V100 and AMD MI250X GPUs. Overall, the MI250X is slightly faster for most kernels but shows more sensitivity to register pressure. We also report scaling results for a distributed memory implementation on up to 16 MPI ranks. 
\end{abstract}


\maketitle

\newpage

\newpage
\newpage
\vfil \break
\clearpage
\section{Introduction} \label{s:intro}

Particle-in-cell (PIC) solvers combine grid-based methods with Lagrangian particle tracking methods and Monte Carlo sampling for collisional processes. In this paper, we focus on PIC solvers for non-equilibrium, {\idef {low-temperature plasmas(LTPs)}} \citep{alves-turner18}. LTPs find applications in semiconductor processing, advanced manufacturing, and  materials science.

LTPs are modeled by a set of time-dependent partial differential equations (PDEs), which include the {\em electron Boltzmann transport equation}: a 6D integro-differential equation for the electron distribution function $f$ as function of  position, velocity, and time~\citep{villani2002review}. LTP simulations can be expensive as they involve multiple solvers and time scales that include nonlinear transport, diffusion, and collisional processes. \ipoint{Collisional processes}  comprise elastic, ionization, excitation, recombination, and electrostatic  interaction kernels each presenting different computational challenges, e.g., creation and destruction of particles, and two- and three-body interactions. The creation, destruction, and electrostatic interactions  of particles introduce irregular data access patterns and race conditions. The irregular data access is most pronounced in the \ipoint{Coulombic collision kernel}, a non-linear collision operator capturing electron-electron electrostatic  interactions. Furthermore, calculating rate coefficients, averages that are functionals of $f$, may require capturing the high-energy tails of $f$, which can require tens of thousands of particles per cell. As common applications require millions of time steps, simulations of LTPs can be quite expensive.

\subsection{Contributions} 

The closest work to this paper is from our group~\citep{almgrenbell-e22}, where, using a standard PIC scheme (Section~\ref{s:physics}), we introduced and analyzed computational kernels for particle computations and their coupling to a grid solver (Section~\ref{s:kernels}). 

Moreover, in ~\cite{almgrenbell-e22}, we also examined the feasibility of Python-based rapid prototyping and performance portable techniques such as Numba~\citep{lam2015numba}, CuPy~\citep{nishino2017cupy}, and PyKokkos~\citep{pykokkos21} (Section~\ref{s:tools}). 

PyKokkos is a Python framework that implements the Kokkos programming model for performance-portable shared memory parallel programming~\citep{TrottETAL21Kokkos}.

In this paper, we extend our initial work in several ways. 

\begin{itemize}
    \item We introduce fast algorithms for electron-electron Coulombic interactions. These interactions have  a more challenging structure as they are non-local interactions that need to be efficiently approximated. From a modeling perspective Coulombic electron-electron interactions become important once the ionization degree exceeds 0.1\%, thus are prevalent in many particle codes. 

    \item Using PyKokkos, we demonstrate performance portability of our implementation and we evaluate in  on an AMD MI250X system and compare its performance with an NVIDIA Volta 100 system. 
\end{itemize}

We also discuss the \emph{null collision} method that it is quite popular in PIC codes and we show that it the plasma regimes we're interested in, the null collision method doesn't yield computational advantages.

 
\section{Mathematical Formulation and Discretization}\label{s:physics}

LTPs include  argon, helium, air, carbon oxides, combustion gases, and others~\citep{alves-turner18,lxcat}. Such plasmas are governed by elastic,  ionization, excitation, and recombination collisions coupled to species transport and possibly a background reacting compressible fluid flow. Depending on the plasma ionization degree, Coulombic interactions between electrons,  can have a major impact \citep{Hagelaar_2016}. To make ideas concrete and present our numerical experiments, we consider an argon LTP with no background flow. This setup is a typical configuration in glow discharge experiments and a good test problem for our solver~\citep{kongpiboolkid2015}. Other types of gases can be substituted, for example helium, nitrogen, or air and the structure  of the equations remains the same, the main difference being the number of species and the number of different collisions.

An argon LTP state is described by the following:
\begin{inparaenum}[(i)]
\item   the \idef{``heavies''}, comprising the \idef{neutral argon} $n_n(x,t)$;
\item the \idef{ionized argon} $n_i(x,t)$, and \idef{metastable argon} $n_m(x,t)$ (particles per unit volume),  which are evolved using transport continuum PDEs;
\item the \idef{free electrons}, which  are treated kinetically with  $f(x,v,t)$ being the \idef{electron distribution function} (in m$^{-6}$/s);
\item   and $\phi(x,t)$ (in volts), the instantaneous \idef{electric potential} driven by a radiofrequency imposed potential and charge imbalances in the plasma.
\end{inparaenum}  
Let $\Omega$ represent the physical domain, so that $x \in \Omega$ and $v \in \mathbb{R}^3$. Then, the  evolution of the state variables $n_n,n_i,n_m,\phi,f$ is given by
\begin{linenomath*}
\begin{subequations}\label{eqn:ltp}
\begin{align} 
  \frac{\partial f}{\partial t} + v\cdot \igrad_x f + q \igrad \phi\cdot \igrad_v f = \sum_{k, c} \mathcal{C}_{kc} [f] + \mathcal{C}_{e}[f]&,  \label{eqn:f}\\
  -\ilap \phi = z (n_i-n_e[f])&,  \label{eqn:phi}\\
    \frac{\partial n_k} {\partial t} - \nabla \cdot (\mu_k n_k \igrad \phi+D_k\nabla n_k) = \sum_c \mathcal{C}_{kc}[f]&.    \label{eqn:n}
\end{align}
\end{subequations}
\end{linenomath*}
Here $k \in \{ i,m,n\}$ for ion (Ar$_+$), metastable (Ar$_m$), and neutral argon atoms; $q,z$ are constants related to vacuum properties, electron mass, and charge; $D_k$ and $\mu_k$ are the corresponding diffusion and mobility constants for each species; $n_e[f] = \int_v f$ is the number of  electrons per unit volume. $\mathcal{C}_{kc}(x,t) = \int_v \sigma_{kc} (\|v\|_2)\|v\|_2 f(x,v,t)$ represent source and sink terms for the fluid equations in Equation~\ref{eqn:n}. 
$\mathcal{C}_{e}$ represents the Coulombic interactions between electrons. $\mathcal{C}_{kc} $ represents the $c_\mathrm{th}$ collision operator~\citep{vahedi1995} between electrons and species $k$ and depends on the \ipoint{collision cross-section} $\sigma_{kc}(\|v\|_2)$. For example, electron-heavy elastic and excitation collisions are given by  $\mathcal{C}_{kc}[f](v,x,t)= n_k(x,t) \int_{s} (f(v'(v),x,t)-f(v,x,t))\sigma_{kc}(\|v\|_2)\|v\|_2 $, where $s$ is the solid angle,  $\sigma_{kc}(\|v\|_2)$ is an experimentally determined, electron-energy dependent \ipoint{collision cross-section}, $v$ is the post-collision velocity,  and $v'$ the pre-collision velocity respectively. The expressions for three body collisions and Coulombic interactions are  more involved and omitted here. In our simulations we include five electron-heavy collisions, summarized in Table~\ref{t:collisions}, which are representative of the collisions found in LTPs. Finally, Equations~\ref{eqn:f}--\ref{eqn:n} are furnished with boundary and initial conditions.

\begin{table}
  \centering
\caption{ Electron-Heavy particle  collisions.\label{t:collisions}}
\begin{tabular}{llr}
\hline
C1:&  $\mathrm{Ar} + e \rightarrow \mathrm{Ar} + e$ & elastic electron-neutrals \\
C2:&  $\mathrm{Ar} + e \rightarrow \mathrm{Ar}_+ + 2e$ & ionization \\
C3:&  $\mathrm{Ar} + e \rightarrow \mathrm{Ar}_m +e$ &  metastable excitation \\
C4:&  $\mathrm{Ar}_m + e \rightarrow \mathrm{Ar}_+ +2e$ & 2-step ionization \\
C5:&  $\mathrm{Ar}_+ + 2e \rightarrow \mathrm{Ar}_m +e$ & recombination \\
\hline\\[4pt]
\end{tabular}
\end{table}

These collisions represent multispecies elastic and inelastic collisions found in most LTPs. C2 and C4 ``create'' electrons by knocking them off neutral argon atoms;  C5 removes free electrons by recombining them with Ar$_{\mathrm{+}}$ to form Ar. The second electron in C5 acts as a ``catalyte'' in this reaction, and therefore  is a three-body collision involving two electrons, which is more computationally expensive and involves special algorithms that do not seem to be discussed in the literature for GPU implementations.

In addition to the standard electron-heavy collisions, we model Coulombic interactions, which are interactions between two charged particles that result from their local electric fields. Coulombic interactions can be electron-electron, ion-ion, and electron-ion interactions. Here, we limit discussion and implementation to electron-electron Coulombic interactions, which have the largest effect on the LTPs we consider. These interactions are frequently modeled as binary elastic collisions in a manner similar to collision C1 but between two electrons.


\subsection{Discretization} \label{s:discretization}
Following state-of-the-art hybrid PIC-formulations~\citep{amrex21}, we discretize Equations~\ref{eqn:f}--\ref{eqn:n} as follows. We assume that $\Omega$ is meshed using a regular grid with \idef{$M$ ``cells''} $\{\omega_j\}_{j=1}^M$; Equations~\ref{eqn:phi} and \ref{eqn:n} can be discretized on the regular grid using any standard method; we use finite-differences. We discretize Equation~\ref{eqn:f} using a \idef{Direct-Simulation-Monte-Carlo scheme (DSMC)}:  a Lagrangian scheme that tracks the motion of $N(t)$ particles with positions $x_l$, velocities $v_l$, and weights $w_l$. These particles represent a $N(t)$-point sample from $f(x,v,t)$: we define $\mathcal{F}(t)=\{x_l,v_l, w_l\}_{l=1}^N$ and thus, $\int_{\omega_j}\int_v f(x,v,t)g(x,v)  \approx \sum_{x_l \in \omega_j} w_l g(x_l,v_l)$, $\forall \omega_j \subset \Omega$, for any function $g(x,v)$.$N(t)$ varies in time due to collisions and boundary conditions. Particle position and velocity updates correspond to advecting $f$ in the left-hand-side of Equation~\ref{eqn:f}; the Monte-Carlo step approximates the right-hand side of Equation~\ref{eqn:f}.  Then, given the \idef{current state} $n^0_k$, $\mathcal{F}^0$, $\phi^0$, and time step $\delta$, we advance to the \idef{new state} $n^+_k$, $\mathcal{F}^+$, $\phi^+$ using the operator-split scheme summarized in Table~\ref{t:pic}. These steps are detailed below, where $\phi$ and $n_k$ are $M$-dimensional grid functions and $F$ are particles sampled from $f$.
 \begin{table}
  \normalsize
\caption{PIC Scheme\label{t:pic}. Index $c$ indicates collision type; index $k$ species; $l$ particle; and $j$ spatial cell.}
\begin{tabular}{|l|r|}
  \hline
  Step S1: & \multirow{2}{*}{$\mathcal{F}^{0,'} = \{x_l^c,v_l^c, w_l^c\} = \mathcal{C}_{e}[\mathcal{F}^0]$} \\
  DSMC-Coul &  \\
  \hline
  Step S2a: & \multirow{2}{*}{$\mathcal{F}^c = \sum_{k,c}\mathcal{C}_{kc}[\mathcal{F}^{0,'}]$}   \\
  DSMC-Coll & \\
  \hline
  Step S2b: & \multirow{2}{*}{$E_l^c= - q \igrad \phi(x^c_l),\ \forall i$} \\
  DSMC-C2P & \\
  \hline
  Step S2c: & \multirow{2}{*}{$\mathcal{F}^+ = \{ x_l^+ = x^c_l + \delta v^c_l;\ v^c_l + \delta E_l^c \}_l$} \\ 
  DSMC-Push & \\
  \hline
  Step S3a: & \multirow{2}{*}{$\forall j, \ n^+_e(\omega_j) = \sum_{x^+_l \in \omega_j} w_l $} \\
  P2C & \\
  \hline
  Step S3b: & \multirow{2}{*}{$\forall j,k,c,$ compute $\mathcal{C}^+_{kc}[\mathcal{F}^c](\omega_j)$} \\
  P2C & \\
  \hline
  Step S4a: & \multirow{2}{*}{$\forall k,\ n_k^+=$  Solve \ref{eqn:n} w/ $\phi^0$ and $\mathcal{C}_k^+$} \\
  PDE & \\
  \hline
  Step S4b: & \multirow{2}{*}{$\phi^+ = $ Solve \ref{eqn:phi} w/ $n^+_i, n^+_e$}  \\
  PDE & \\
  \hline
\end{tabular}
\end{table}

\begin{itemize}
    \item Step S1 is the Coulombic collision step. Discussion can be found in Section~\ref{s:coul} and the algorithm is detailed in Table~\ref{t:coulkernel}.
    \item Step S2a models the remaining collisions, with the discussion of recombination (a 3-body collision) separated
        \begin{itemize}
            \item The recombination algorithm is discussed in Section~\ref{s:reco} and detailed in Table~\ref{t:recomb}.
            \item The remaining collisions are discussed in Section~\ref{sec:collision:kernel} and detailed in Table~\ref{t:coll_steps}.
        \end{itemize}
    \item Steps S2b and S2c represent the cell-to-particle interpolation of the electric field and the application of advection in physical and velocity space. This step is discussed in Section~\ref{sec:collision:kernel}.
    \item Step S3 are \idef{particle-to-cell operations (P2C)} used for calculating $n_e(x)$, $T_e(x)$, and the right-hand side of Equation~\ref{eqn:n}. This step is discussed in Section~\ref{s:p2c}.
\end{itemize}

Steps S4 are PDE solves that require $\bigO(M \log M)$ work and $\bigO(\log M)$ depth, assuming that optimal algorithms like multigrid are used for Equation~\ref{eqn:phi}. \emph{Fast solvers for  steps S4 are not in the scope of the paper as they have been studied extensively in the literature}~\citep{amrex21}. These solvers are crucial in 3D3V (3D in space, 3D in velocity), but in 1D3V and 2D3V (1D and 2D in space, respectively; 3D in velocity), the main cost is the Boltzmann solver and translation between the particle and cell discretizations~\citep{alves-turner18}. Here we use explicit second-order finite differences with upwinding for the transport equations; a standard-second order stencil; a sparse Cholesky solver for the Poisson problem; and focus on the 1D3V case for the PDEs. But in the DSMC and P2C parts, we assume 3D3V so that our algorithms correspond to that of 6D simulations with regular 3D grids.  For irregular 3D grids, the only change is that sorting particles to cells requires a binary search.



We briefly discuss the cost of the scheme and revisit it in Section~\ref{s:results:perf}.  As we will see, step S1 involves forming particle pairs by cell; our implementation has $\bigO(N)$ work and $\bigO(1)$ depth. Ignoring particle creation and destruction (Section~\ref{sec:collision:kernel}), steps S2 have  $\bigO(N)$ work and $\bigO(1)$ depth (for regular grids). Steps S3 require sorting particles to cells followed by segmented reductions, thus, requiring $\bigO(N+M)(\log N)$ work and $\bigO(\log N + \log M)$ depth~\citep{blelloch1996}. An alternative implementation uses atomic operations for the reductions while avoiding sorting. We discuss it in Section~\ref{s:p2c}.


\MyPara{Summary}
Using a well-known PIC scheme, we focus on step S1 (Coulombic kernel), steps S2a--S2c (Boltzmann kernel) and steps S3a and S3b for the P2C calculations. Optimizing these steps is of extreme importance as they constitute the bulk of the computations in most LTP simulations.  We detail these kernels in Section~\ref{s:kernels}.

\section{HPC Productivity in Python} \label{s:tools}

In this section, we discuss the Python-based HPC tools we used in our PIC code. Python has seen increasing use in the field of scientific computing ~\citep{Oliphant07ScientificPython,ziogas2021productivity,bauer2019code}. Our code uses several well established  libraries: NumPy~\citep{HarrisETAL20Numpy}, CuPy~\citep{nishino2017cupy}, and SciPy~\citep{VirtanenETAL20Scipy}.  We also used Numba~\citep{lam2015numba}, a just-in-time compiler for Python that compiles handwritten Python kernels for CPUs or NVIDIA GPUs. However, the Numba support for AMD GPUs appears to be inactive.

In our group, we have developed an alternative to Numba, \textbf{PyKokkos} \citep{pykokkos21}. It provides portable Python abstractions for \ipoint{performance-portable} shared memory parallel programming of custom kernels. PyKokkos then translates kernels written by the user into Kokkos~\citep{TrottETAL21Kokkos} and C++, using \href{https://pybind11.readthedocs.io/en/stable/basics.html}{pybind11} for interoperability between the two languages. PyKokkos memoizes the compiled functions to avoid the cost of re-compiling the kernel after the first call. PyKokkos through Kokkos uses OpenMP for CPUs, CUDA for NVIDIA GPUs, and HIP for AMD GPUs. Due to the performance portability abstractions, we do not need to change our code in the future to run it on these devices. We implemented our GPU kernels using PyKokkos: for every Numba or CUDA GPU kernel, we also have a PyKokkos implementation.

\section{Particle Kernels}  \label{s:kernels} 

In this section, we describe the data structures we used (Section~\ref{sec:data:structures}),  details regarding algorithms, implementation, and performance for the electron-neutral collision kernel (Section~\ref{sec:collision:kernel}), the recombination collision C5 in Section~\ref{t:collisions},  (Section~\ref{sec:recombination:kernel}), the particle-to-cell kernels (Section~\ref{s:p2c}), and the Coulombic interactions (Section~\ref{s:coul}).

\subsection{Data Structures}
\label{sec:data:structures}

We store all  particle positions and velocities together in a
two-dimensional $N$-by-$6$ CuPy array. CuPy arrays are stored in row major order to improve locality during the DSMC collision kernel, as no two particles need to communicate within that kernel.
Particle weights, source term contributions, and energies--- are stored together in a separate two-dimensional $N$-by-$4$ CuPy array. Typically, the particle number will grow due to ionization until the system reaches steady state. During setup, we preallocate extra  memory buffers during the setup phase for all data structures; the size of these buffers is estimated analytically. For example, if we expected a 10x growth in particle number during a simulation, we would set the total size of the array to $N_{\mathrm{max}} = 10 N_{\mathrm{init}}$. In this way, we avoid dynamic memory allocation.


\subsection{DSMC Collision Kernel}\label{sec:collision:kernel}

\begin{table}
  \centering
  \normalsize
\caption{DSMC Collision Step for a particle $l$.\label{t:coll_steps}}
\begin{tabular}{lr}
\hline
Draw six random numbers $R_l$ per particle & CS0 \\
Load to registers particle data $x_l,v_l,w_l$ & CS1\\
$\forall k,c$ compute cross-sections $\sigma_{ck}(||v_l||_2)$ & CS2\\
$\forall k,c$ compute probabilities $\pi_{lck}(\sigma_{ck})$ & CS3\\
Collision: update $x_l,v_l$ & CS4\\
Collision: if ionization, create particle $l': x_{l'},v_{l'},w_{l'}$ &\ CS5\\
Advection: update $x_l,v_l$ & CS6\\
Boundary conditions: if $x_l \notin \Omega$ from advection, set $w_l = 0$ & CS7\\ 
\hline\\[4pt]
\end{tabular}
\end{table}

We now discuss the details of the DSMC collision kernel, which
performs steps S2a-c (Table~\ref{t:pic}) of the PIC scheme. This step is
not embarrassingly parallel due to the recombination collision C5,
which we discuss separately in Section~\ref{sec:recombination:kernel}, and
due to particle creation (e.g., ionization) or loss (e.g., absorption
by electrodes or dielectric walls, or recombination). We verify the correctness of the DSMC collision kernel against BOLSIG+, an industry-standard electron Boltzmann solver \citep{Hagelaar_2005}. Despite the inherent randomness to this step, results are repeatable with negligible variance. The details of this comparison are beyond the scope of this paper.

\MyPara{Algorithm}
Table~\ref{t:coll_steps} details the steps. In the \ipoint{DSMC-Coll step},
for each particle $l$ and each collision $c$ with species $n_k$, we
compute collision probabilities $\pi_{lck}=1-\exp(-\Delta t \|v_l\|_2 \sigma_{ck}(\|v_l\|_2) n_k)$.
Using $\pi_{lck}$, we decide whether and how to collide
particle $l$, using (six in our case) random numbers. For each timestep, we precompute 
these random numbers for all particles using \CodeIn{cupy.random.rand()} before the kernel invocation. If the particle collides, we compute the post-collision velocities using conservation
of mass and energy~\citep{vahedi1995}. Last, advection in physical and
velocity space is applied and boundary conditions are enforced.
Referring to Table~\ref{t:pic}, notice that before CS6 we need the
\ipoint{C2P operation} to compute $\igrad \phi(x_l)$. Depending on the
scheme, this can be an expensive operation. Most PIC-DSMC
codes are low-order accurate and use constant interpolation; we do the
same. We precompute a constant $\igrad \phi$ per cell, and then retrieve it using the particle's cell ID. 

Accuracy and numerical stability
dictate overall collision frequency to be low, say less than $10\%$. 
As a result, the kernel spends most of its time in
determining which collision (if any) a particle will undergo (steps
CS3-4 in Table~\ref{t:coll_steps}), rather than performing collision
calculations (step CS6). To determine $\pi_{lck}$, all
particles need to compute $\sigma_{ck}(\|v_l\|_2)$, a function of the 
particle's speed; typically, this is done via lookup tables \citep{lxcat}. Instead, we predetermine a piecewise polynomial interpolation of each cross section. By doing so, we reduce calculation time while keeping cross section error under $2\%$.

A potential efficiency pitfall is thread divergence due to particles
undergoing different collisions. However, due to low collision rates, most
particles undergo no collisions within each time step. Alternative implementations designed to minimize thread divergence proved to be more costly due to the extra calculations required, which outweighed the minimal thread divergence.

As mentioned, a challenge in implementing the collision kernel is the varying number of particles in each time step. Before executing the collision kernel, it is unknown how this number will change due to creation or destruction of particles. To address this problem, we maintain a single write index $I$ shared by all particles. When a particle is created, each thread increments $I$ \emph{atomically} and writes the new particle into the array at that index; thus, a full size $N$ buffer is not required. The other alternative considered was the use of buffers giving particles unique write indices for potential particle creation. We found that the cost of data movement and reshuffling of particles in the non-atomic algorithm was prohibitively expensive. Given the support for atomic operations in modern GPU architectures along with relatively low ionization rates, the additional cost of using atomic operations is minimal. We observed up to a 6$\times$ speedup of this kernel when using atomics.

\begin{figure}
  \centering
  \includegraphics[width=0.5\textwidth]{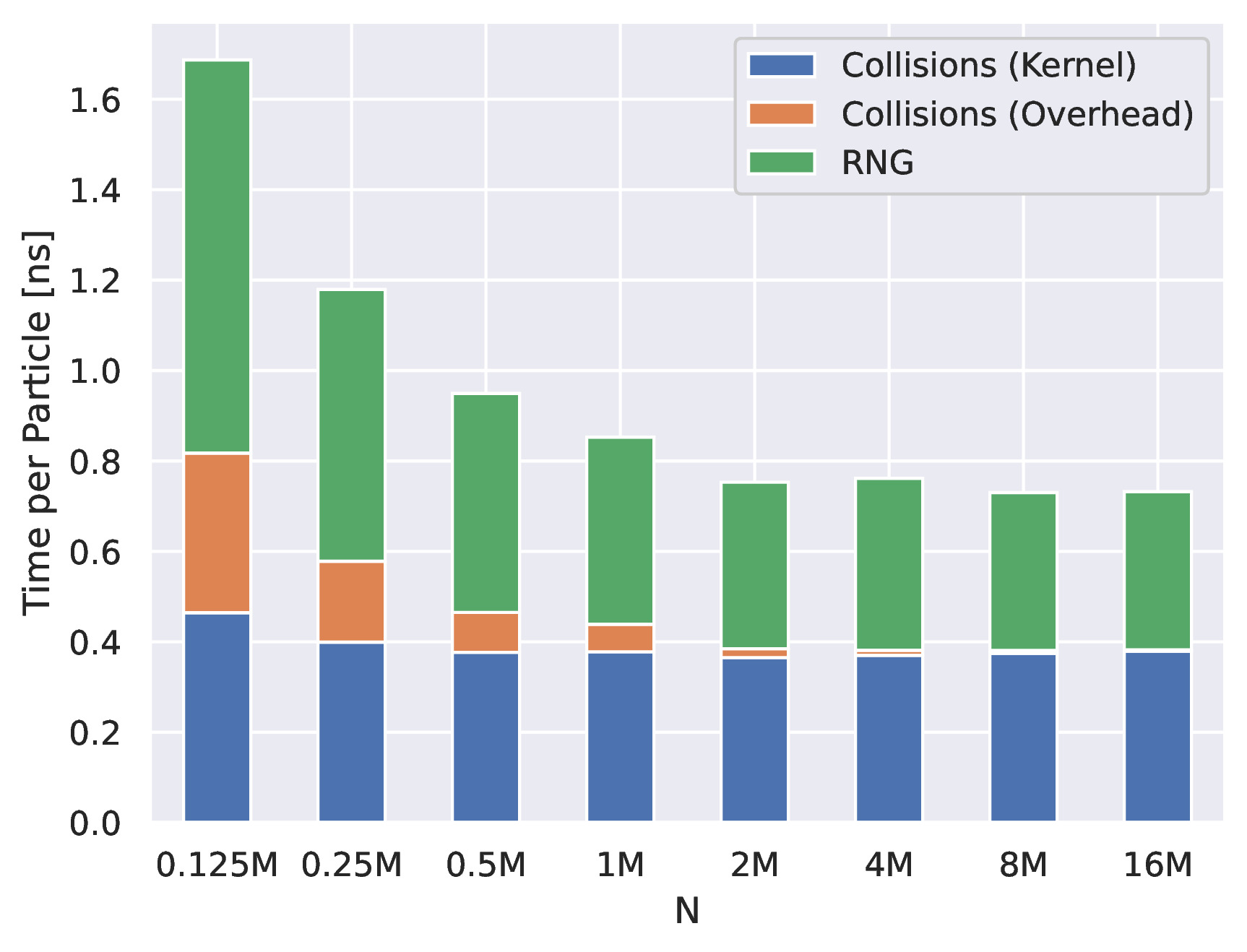}
  \vspace{-8pt}
    \caption{DSMC Collision Kernel Breakdown: PyKokkos Atomic Updates.}
    \label{fig_ek_atomic}
\end{figure}

\MyPara{Performance}
We implemented the collision steps with PyKokkos kernels.
Figure~\ref{fig_ek_atomic} shows the breakdown of the execution time of the particle steps for the PyKokkos implementation of the atomic version of the DSMC collision kernel. We split the runtime into kernel time (measured with the NVIDIA profiler \CodeIn{nvprof}) and overhead time (wall clock time minus kernel time); we also show the RNG time. 


\MyPara{Operator Splitting Algorithms}
We note that implementing particle creation using atomics allows for a simple extension to using operator split methods, where multiple collision kernels are performed for each time step of the continuum physics due to different time step constraints. The use of atomics
removes the need for a synchronization step after each call to the collision kernel, as there is no potential for overwriting particle data. Further, we are able to run multiple time steps within one call to our DSMC collision kernel, removing overhead costs of relaunching the kernel. The use of operator splitting is further beneficial when considering communication costs in multi-GPU implementations, as demonstrated in Figure~\ref{f:mpi}. As further discussed in Section~\ref{s:results:mpi}, we consider a simple MPI implementation with domain replication.

\MyPara{Null Collision Method} To evaluate the performance of our collision kernel, we compare our method to an alternative collision selection method frequently used in the literature, the null collision method \citep{vahedi1995}. Instead of each particle individually computing a collision probability at each timestep, a "null probability", labeled $P_{\mathrm{NULL}}$, is computed initially to be the maximum possible particle collision probability. This value is typically on the order of $10^{-2}$. Each timestep, only $P_{\mathrm{NULL}} N$ particles are considered for collision. This method is designed such that the same number of total collisions occur, but fewer particles must be considered for collision. By reducing the number of particles that are considered for collision at each timestep, the null collision method reduces the number of floating point computations. In theory, this should improve overall performance, most notably by eliminating often-expensive collision cross section computations. When we implemented this method in our code, we found that the null collision method actually slowed our execution time substantially. With the null method, the reduction in floating point operations performed in the collision kernel resulted in minimal runtime improvements. In both cases, all particle data must still be loaded by the kernel for step CS6; no memory operations are saved. In fact, we observe that the time taken to randomly select the set of $P_{\mathrm{NULL}}$ particles to be tested for collision at each step causes an overall drop in performance of the kernel. Because we find that with our implementations the null collision method performs worse, we choose to stick with the original collision selection method.

\subsection{Kernel for Recombination Collision C5}\label{s:reco}
\label{sec:recombination:kernel}

\newcommand{\pli}{\ensuremath{\mathcal{P}}}
\newcommand{\cli}{\ensuremath{\mathcal{P}_j^c}}

\begin{table}
    \vspace{10pt}
  \centering
  \normalsize
  \caption{Recombination Steps. RS0 is precomputed; RS1 is a separate GPU kernel; RS2--RS5 are in the same GPU kernel.
  \label{t:recomb}}
    \begin{tabular}{lr}
    \hline
    CS3~(in \ref{t:coll_steps}) computes  $\pli$  & RS0 \\
    $\forall$ cell $\omega_j$, construct $\cli$ & RS1 \\
    $\forall$ particle $l \in \pli:$ & \\
    \hspace{10pt} Read $\omega_j(l)$ & RS2 \\
    \hspace{10pt} Atomically pop $l' \in \cli(\omega_j(l))$ & RS3 \\
    \hspace{10pt} Update catalyte velocity $v_{l'}$ & RS4 \\
    \hspace{10pt} Destroy primary: $w_l = 0$ & RS5 \\
    \hline\\[4pt]
    \end{tabular}
\end{table}

Collision C5 in Table~\ref{t:collisions} is a three body interaction between Ar$_{+}$ and two electrons, \idef{the primary and the catalyte}. This collision requires global synchronization across GPU-blocks to set up primary-catalyte pairs. This kernel is a component of step S2a from Table~\ref{t:pic}.

\MyPara{Algorithm} C5 is summarized in Table~\ref{t:recomb}. The catalyte particle must be in the same  cell $\omega_j$ as its primary. Each particle in $\omega_j$ is exclusively designated either as catalyte or primary. Furthermore,  primaries must be \ipoint{uniquely} matched to catalytes. We do this as follows. We  define $\pli$ to be the list of all primaries, precomputed in step CS3 (Table~\ref{t:coll_steps}) of the collision kernel. For each primary $l$  we also have precomputed its cell  $\omega_j(l)$.  For each cell $\omega_j$,  we define $\cli$ to be the \idef{set of all possible catalyte particles $l'$ in $\omega_j$} such that $l' \not\in \pli$. Then for a particle $l \in \pli$ we execute steps RS2--RS5 in Table~\ref{t:recomb}.

\MyPara{Implementation} We now consider steps RS1 and RS3, the two major steps of this kernel. One potential implementation of these steps is locally in space; if particle data is sorted by position, these steps can be done by mapping of primaries to non-primaries within a given cell. Since our particle data is unsorted, we instead incorporate atomics in the same manner as the collision kernel. Step RS1 atomically increments a counter for each cell $\omega_j$, such that potential catalyte
partners in cell $\omega_j$ are written into $\cli[I_j]$. Step RS2 is a load operation, as the $\omega_j$ is precomputed. Similar to step RS1, step RS3 reads values from $\cli$ using atomically updated indices to ensure that no two primaries in cell $\omega_j$ select the same catalyte. Steps RS4 and RS5 can be done with minimal computation. We group the steps such that steps RS1 and steps RS2--RS5 are implemented as two separate kernels in both Numba and PyKokkos. 

\MyPara{Performance Consideration} In the physical regime of the glow discharge problem we consider, recombination collisions are rare; thus, the cost for the collision steps RS2-RS5 is relatively low. The majority of time spent in this kernel is therefore in step RS1, the assembly of the catalyte list, which is done for all grid cells. We report detailed timings for this kernel in Section~\ref{s:results}, where we see that the performance of the kernel depends on the number of cells and the number of particles per cell. There we report aggregate timings for steps RS1--RS5. 

\subsection{Coulombic Collision Kernel}\label{s:coul}

The Coulombic collision kernel performs step S1 of Table~\ref{t:pic}. Coulombic interactions are the binary elastic collisions between two charged particles (in this case electrons) that occur as a result of their own electric fields. Each collision exchanges a small amount of momentum between two particles; the Coulombic interactions act as a diffusive process for our electron velocity distribution. Unlike the other collisions, Coulombic collision cross sections are not gathered from data; rather, they are derived from theory \citep{TAoriginal}. 

Similar to the recombination kernel, we require a structure to handle each electron-electron collision; this involves a global synchronization across GPU-blocks. The lack of need to explicitly define primaries and catalytes simplifies our algorithm; the only computational complexity of this kernel comes from the assignment of collisions pairs. Electric fields generated by the electrons are relatively weak, so we only consider collisions between two particles in the same grid cell. Unlike recombination, Coulombic collisions must be modeled at each timestep for \emph{all} particles. We discuss the algorithmic and implementation details below. Our algorithm complexity is linear with respect to number of particles, however it uses atomic operations and irregular data accesses.

\begin{table}
  \centering
  \normalsize
  \caption{Coulombic Collision Steps.\label{t:coulkernel}}
    \begin{tabular}{lr}
    \hline
    $\forall$ cell $\omega_j$, define $P_j =$ \{particles $l$: $x_l \in \omega_j$\} &  \\
    $\forall$ cell $\omega_j$, define $I_j$ to be the starting write index &  \\
    Define $P$, a length $N$ array to store particle indices & \\
    \textbf{CCS1}: Retrieve $N_j = |P_j|$, such that $\sum_j N_j = N$ &  \\
    \textbf{CCS2}: Compute initial write indices $I_j$ using prefix \\ sum over $N_j$ &  \\
    \textbf{CCS3}: Construction of $P$: $\forall$ particle $l$ &  \\
    \hspace{10pt} Determine $j$ such that $x_l \in \omega_j$ &  \\
    \hspace{10pt} Write index of particle $l$ into $P[I_j]$ &  \\
    \hspace{10pt} Atomically increment $I_j += 1$ &  \\
    Define partner pairs $\{l,l'\}$ as $(P[1],P[2]),$ \ldots, \\$(P[N-1],P[N])$ &  \\
    \textbf{CCS4}: Generate two random numbers for each \\
    collision pair $\{l,l'\}$\\
    \textbf{CCS5}: $\forall$ partner pair $\{l,l'\}$ & \\
    \hspace{10pt} Perform collision calculations using $R^{l,l'}$ &  \\
    \hspace{10pt} Update particle velocities $v_l, v_{l'}$ &  \\
    \hline\\[4pt]
    \end{tabular}
\end{table}

\MyPara{Algorithm} We model these Coulombic interactions as binary elastic collisions using the method first presented in 1977 by \cite{TAoriginal} \citep{TAcomparison}. The algorithm is summarized in Table~\ref{t:coulkernel}.

First, we define $P_j$ to be the list of all particles in cell $\omega_j$. Because we store the particles in an unsorted manner, each $P_j$ must be computed as a part of this kernel. Define $N_j$ to be the size of each $P_j$, such that $\sum_j N_j = N$; these values are drawn trivially from the results of the P2C kernel as step CCS1. Rather than move data, we simply track and sort particle indices. Define an array $P$ of length $N$ in which we will store all of the $P_j$. 

In step CCS2, we define initial write indices $I_j$ as the result of a prefix sum over $N_j$. Step CCS3 is done in a single kernel. We construct $\{P_j\}$ by looping over all particles $l$ and computing $j$ such that $x_l \in \omega_j$. The index of particle $l$ is written into $P[I_j]$ and $I_j$ is incremented by 1 atomically. Because we know we will have $\frac{N}{2}$ collision pairs, the collision random number generation (step CCS4) is handled in advance. Similar to the DSMC-Coll kernel, these random numbers are used for generating collisional scattering angles. In our second kernel we implement CCS5, which loops over the list $P$ and performs the desired electron-electron collision computations. Odd number of particles per grid cell are handled trivially.

\subsection{Particle-to-Cell}\label{s:p2c}

The particle-to-cell kernel performs steps S3a-S3b in Table~\ref{t:pic}. It
computes per-cell right-hand sides  for the PDEs in Equation~\ref{eqn:ltp}, as well as quantities of interest, for example the electron
temperature
$T_e(\omega_x)=1/|\omega_x|\int_{\omega_x} \mathbb{E}_f[\|v\|^2]$
\citep{villani2002review}. For our solver, we compute four variables, which are used for the calculation of $n_e(x)$, $n_i(x)$, $n_m(x)$, and $T_e(x)$. For each particle $l$, we define  $\mathbf{V}_l$ to be the vector of these variables. We seek to calculate the sum of these values $\mathbf{V}^j$ corresponding to particles in each grid cell $\omega_j$. That is, $\mathbf{V}^j = \sum_{x_l \in \omega_j} \mathbf{V}_l$.
Therefore, our problem becomes a block reduction, where the prefix of a particle is the spatial cell in which it lies.

\MyPara{Algorithm}
A naive implementation would be to sort the particles to per cell arrays and then use per cell parallel reductions; we found the cost of sorting the particles to be prohibitively expensive. Instead, we use atomic reductions for each $V^j$. This scheme avoids the need for particle sorting. If the number of particles per cell is very large it pays off to introduce further partitioning of $\omega_j$ to auxiliary subdomains $\omega_{jm}$ to atomic updates congestion. Then, with each particle being assigned to a GPU thread, we find $\omega_{jm}(x_l)$,  and  then atomically update $V^{jm}$.  Upon completion a number of threads are assigned to $\omega_j$  for the reduction  $V^j = \sum_m V^{jm}$.

We implemented this atomics-based algorithm using both CUDA and PyKokkos. For $N$ from $100k$ to $16M$, both CUDA and PyKokkos implementations are 10$\times$--20$\times$ faster than our implementation of the non-atomic algorithm. Further performance details are discussed in previous work~\citep{almgrenbell-e22}.

\subsection{Multi-GPU extension} \label{s:results:mpi}

To give a flavor of the relative costs of the solver components in the
context of large-scale applications, we implemented a straightforward
MPI extension: the PDE-calculations (steps S4a and S4b in Table~\ref{t:pic})
are \ipoint{replicated} on each MPI task. For 1D3V and 2D3V
simulations this replication is not prohibitively expensive. In the next section, we show scaling results with up to 160K cells, which is a typical resolution for 2D3V calculations~\citep{hurplasma_gpu19}. Our scheme could be extended successfully to 3D3V calculations in which $\Omega$ is coarsely partitioned in large domains, and each large domain has its own MPI communicator with PDE-grid replication. Our scheme is asymptotically suboptimal compared to domain-decomposing both $\Omega$ and particles, but it is simple to implement and potentially faster for smaller problems. Further analysis is beyond the scope of this paper.

Algorithmically, our MPI extension is a single \CodeIn{MPI\_AllReduce()}. Specifically, let $P$ be the number of \idef{MPI tasks}; each task owns $N/P$ electrons and replicates all $\{\omega_j\}_{j=1}^M$ cells. Steps S1-S2 in Table~\ref{t:pic} are done in an embarrassingly parallel manner at each task. Steps S3a and S3b are where the communication takes place: a local P2C operation (Section~\ref{s:p2c}), to compute a per task $\mathcal{C}^+_k, n^+_k $ followed by an all-reduce with message size $\bigO(M)$ and complexity $\bigO(M \log P)$ \citep{karypis03}. Finally steps S4 are replicated at each MPI task. Since $M=N/\rho $, the $\bigO(N/P)$ calculation dominates when $\rho\gg P$. All tasks are statistically equivalent as they view the entire domain $\Omega$, which eliminates load balancing issues. The overall memory costs are  $\bigO(N/P +  M)$, where the prefactor is 25 and accounts for positions velocities, cross-sections, time-stepping and particle-to-cell interactions.

\subsection{Avoiding Bias} 

Here, we briefly discuss the potential biases of our MPI extension. Our algorithm simulates $\frac{N}{P}$ particles independently on each task. Each GPU runs the DSMC-Coll, Coulombic, and recombination kernels with no communication. Because of the lack of communication here, we only consider Coulombic and recombination collisions between two particles stored on the same GPU. This has the potential to introduce bias in our collision partner selection for these collision types, and thus our system as a whole. However, this bias is inherently negated by our initial distribution of particles. We initialize the particles on each GPU such that each GPU has the same electron probability distribution with respect to both space and velocity. In this way, we consider our overall simulation as $P$ simulations of $\frac{N}{P}$ particles rather than one simulation of $N$ particles. Improved accuracy for our results comes from the extra samples taken not only for $f(v,x,t)$ but also for the quantities $n_e$, $T_e$, $\phi$ which can be calculated with greater accuracy. We experimentally verify that simulation results are nearly identical regardless of how many MPI processes are used. However, further detail regarding self-convergence of PIC methods is beyond the scope of this paper.  
\begin{figure*}
\centering

  \subcaptionbox{Strong; $\rho=1000, N=16e6$.\label{f:r1000s}}{\includegraphics[width=0.25\textwidth]{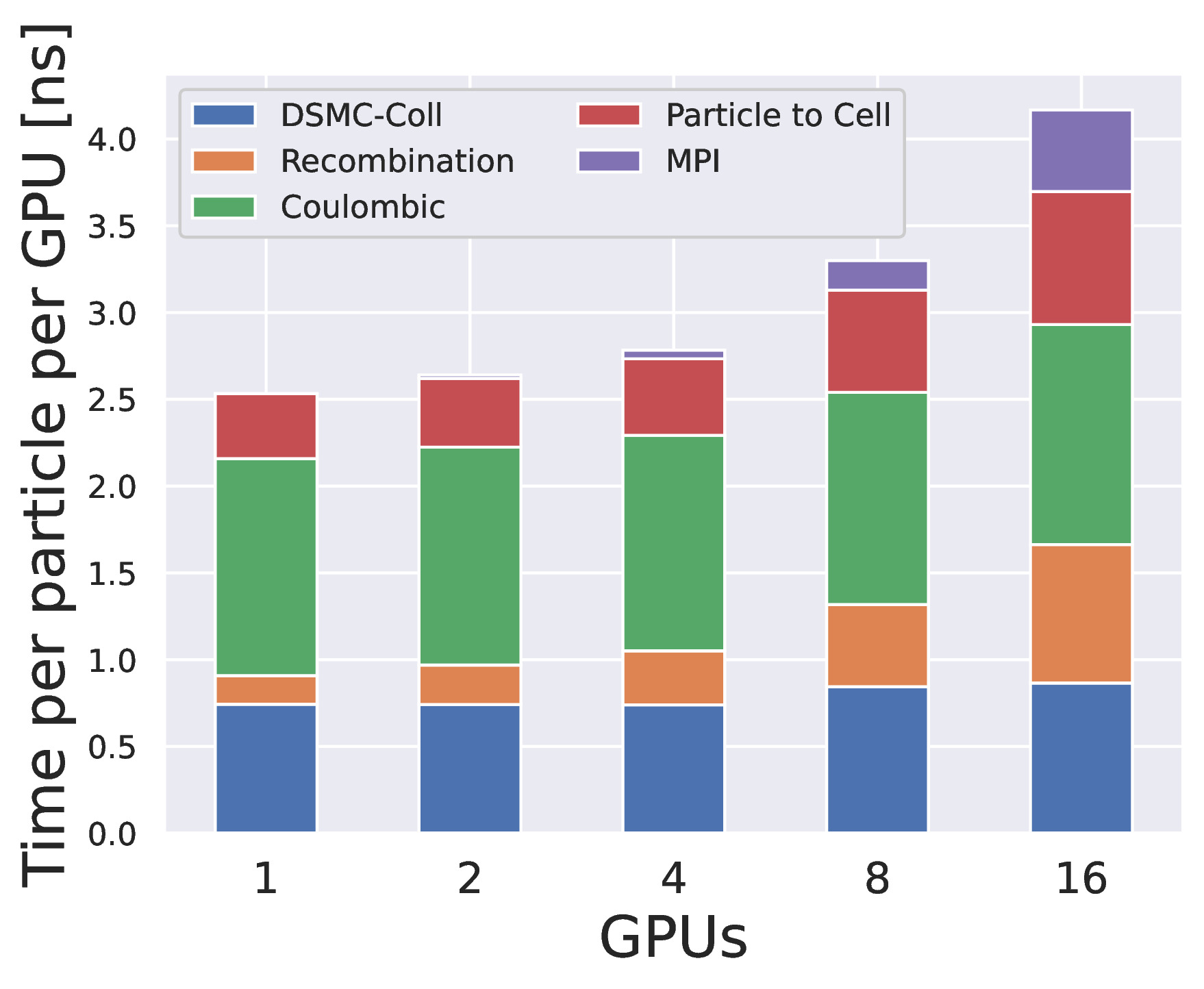}}%
  \subcaptionbox{Weak; $\rho = 1000, n_P = 4e6$.\label{f:r1000w}}{\includegraphics[width=0.25\textwidth]{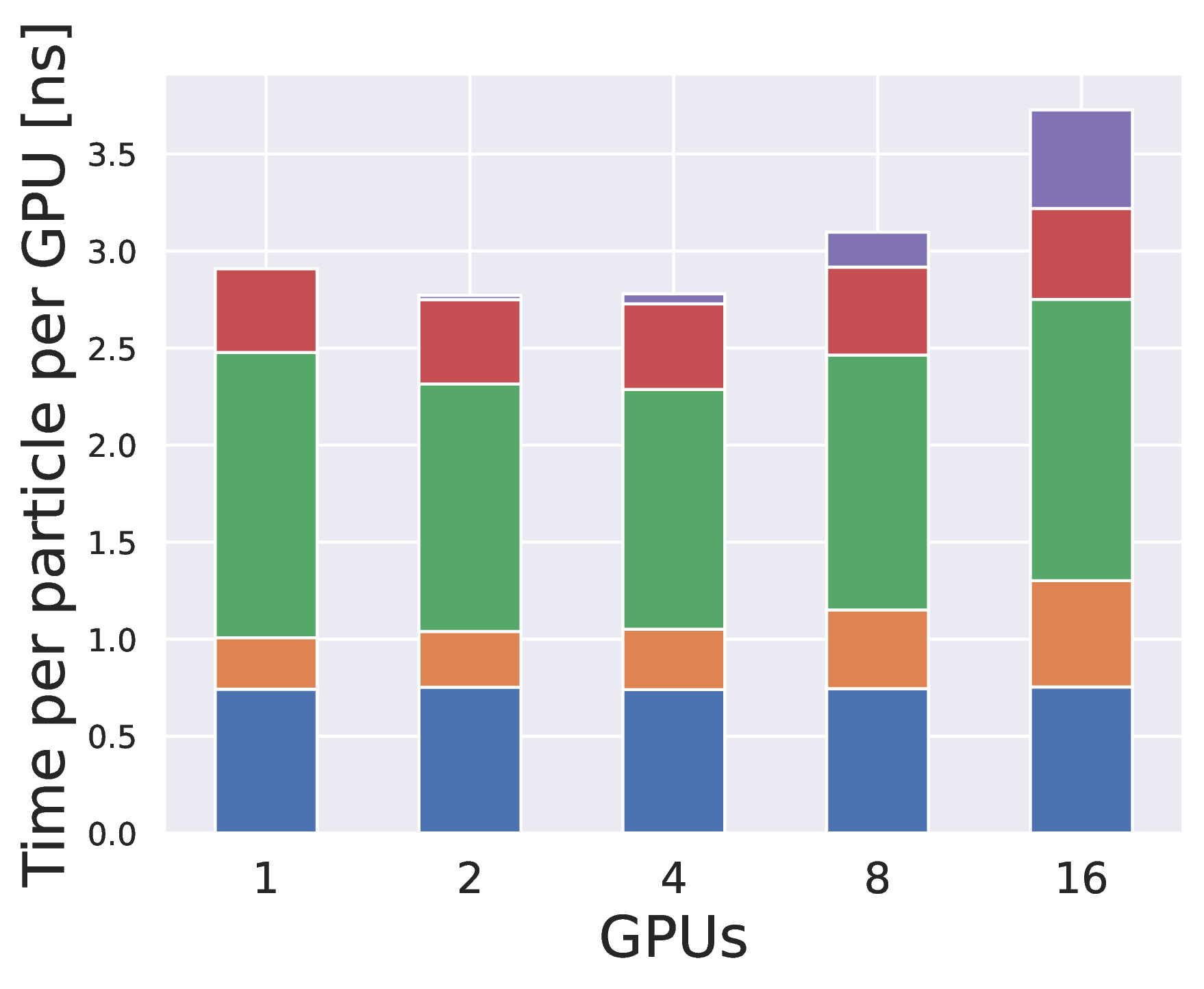}}%
  \subcaptionbox{Strong; $\rho = 100, N = 16e6$.\label{f:r100s}}{\includegraphics[width=0.25\textwidth]{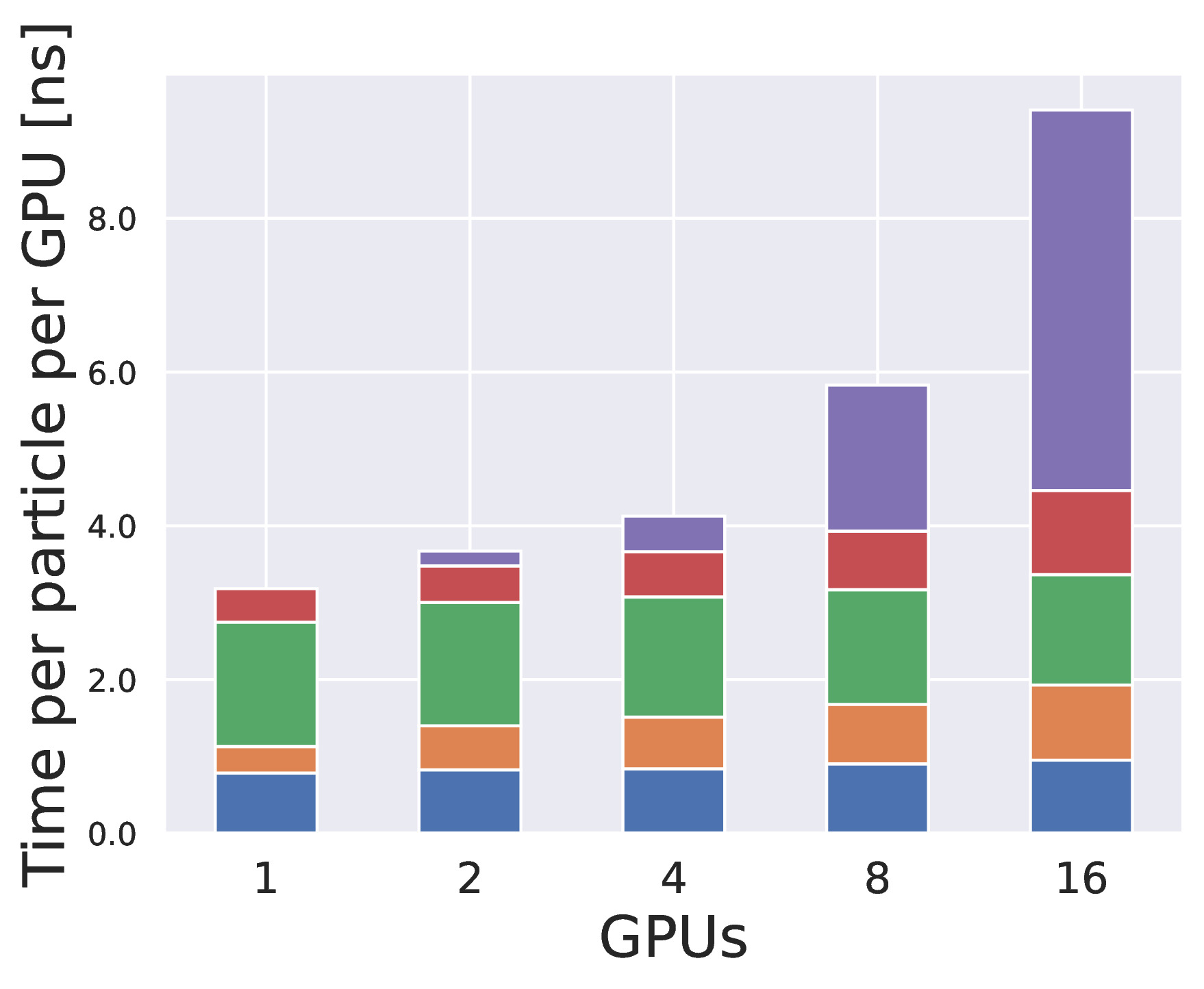}}%
  \subcaptionbox{Weak; $\rho = 100, n_P =   4e6$.\label{f:r100w}}{\includegraphics[width=0.25\textwidth]{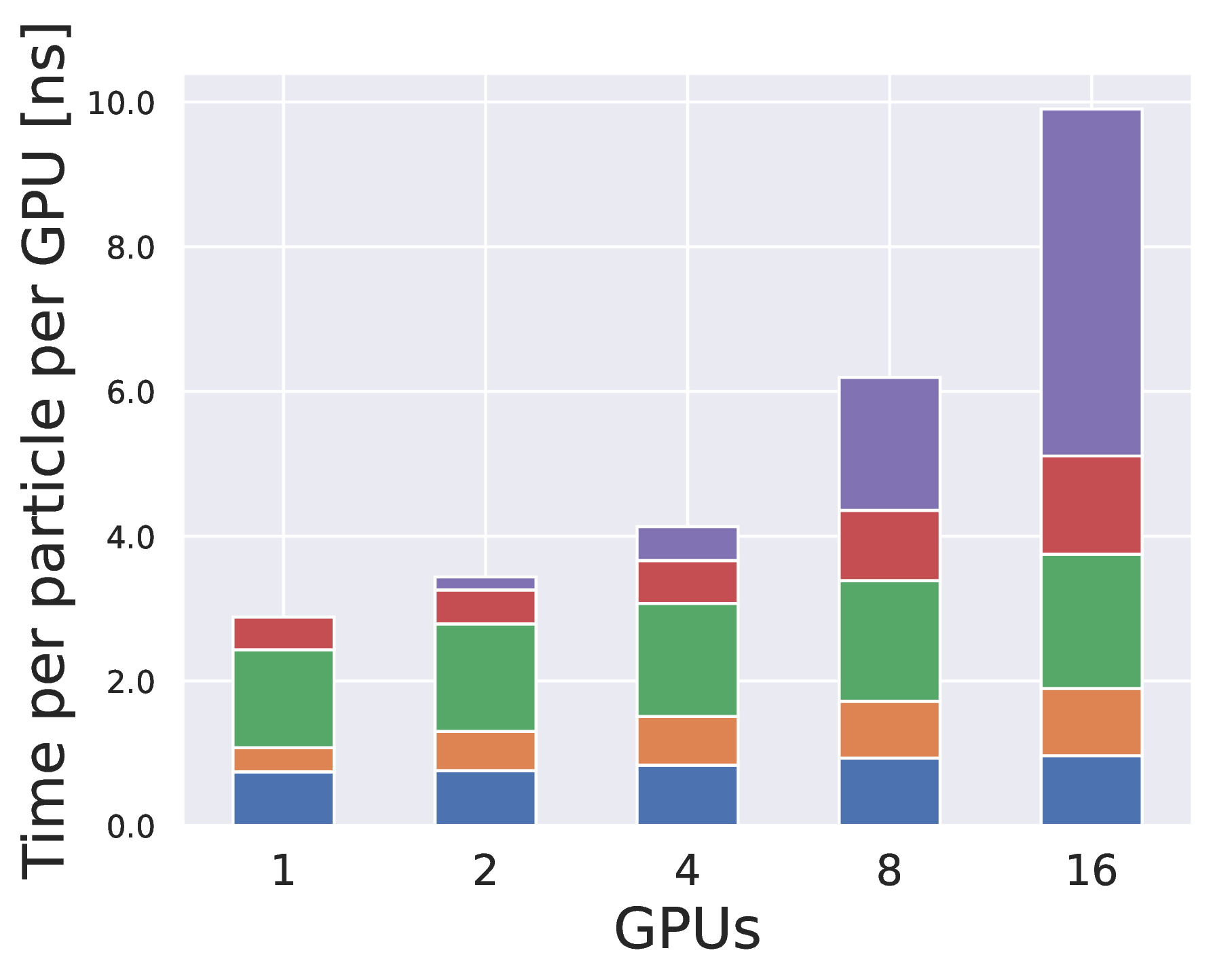}}
  \subcaptionbox{Strong; $\rho = 1000, N = 16e6$.\label{f:r1000sa}}{\includegraphics[width=0.25\textwidth]{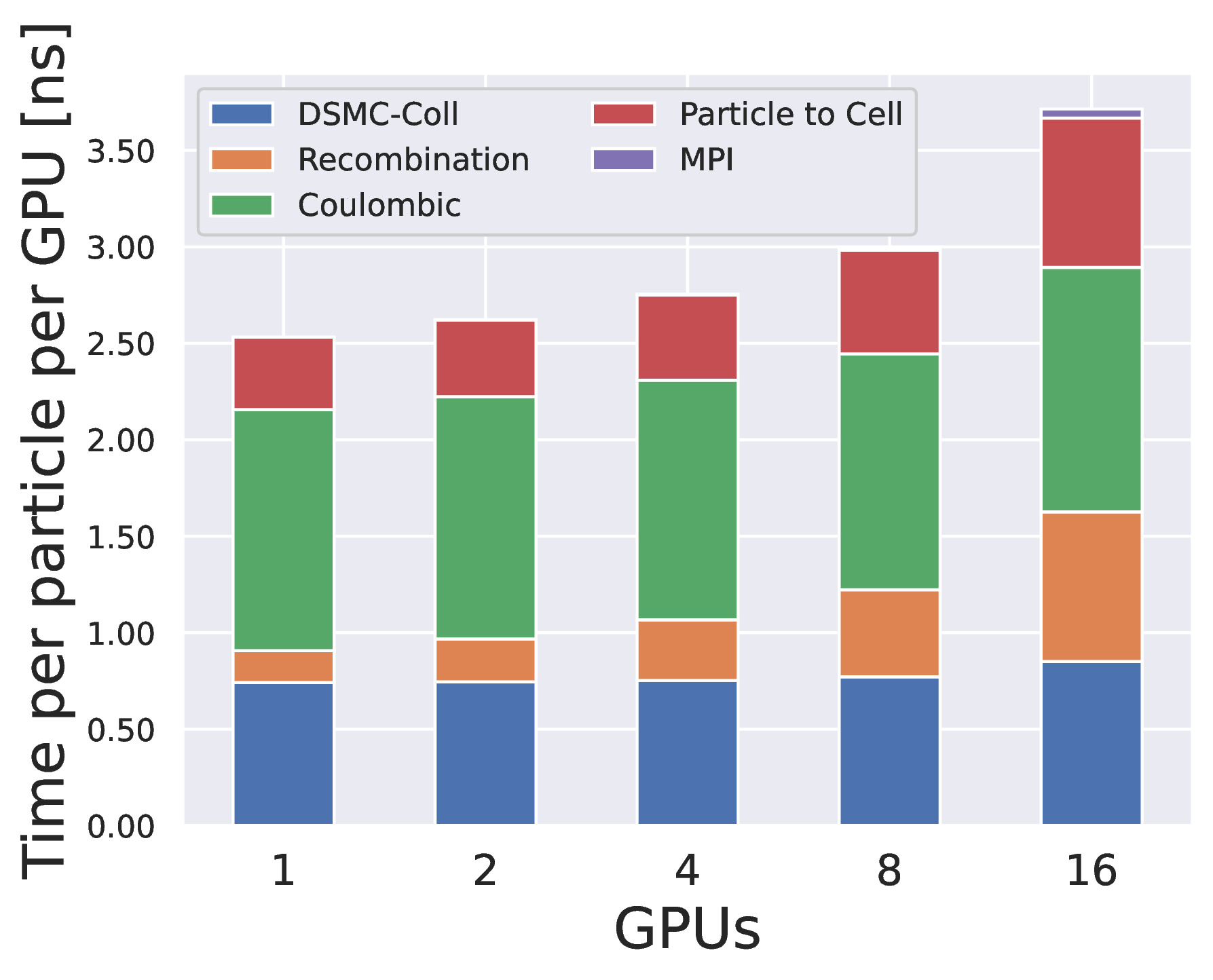}}%
  \subcaptionbox{Weak; $\rho = 1000, n_P =  4e6$.\label{f:r1000wa}}{\includegraphics[width=0.25\textwidth]{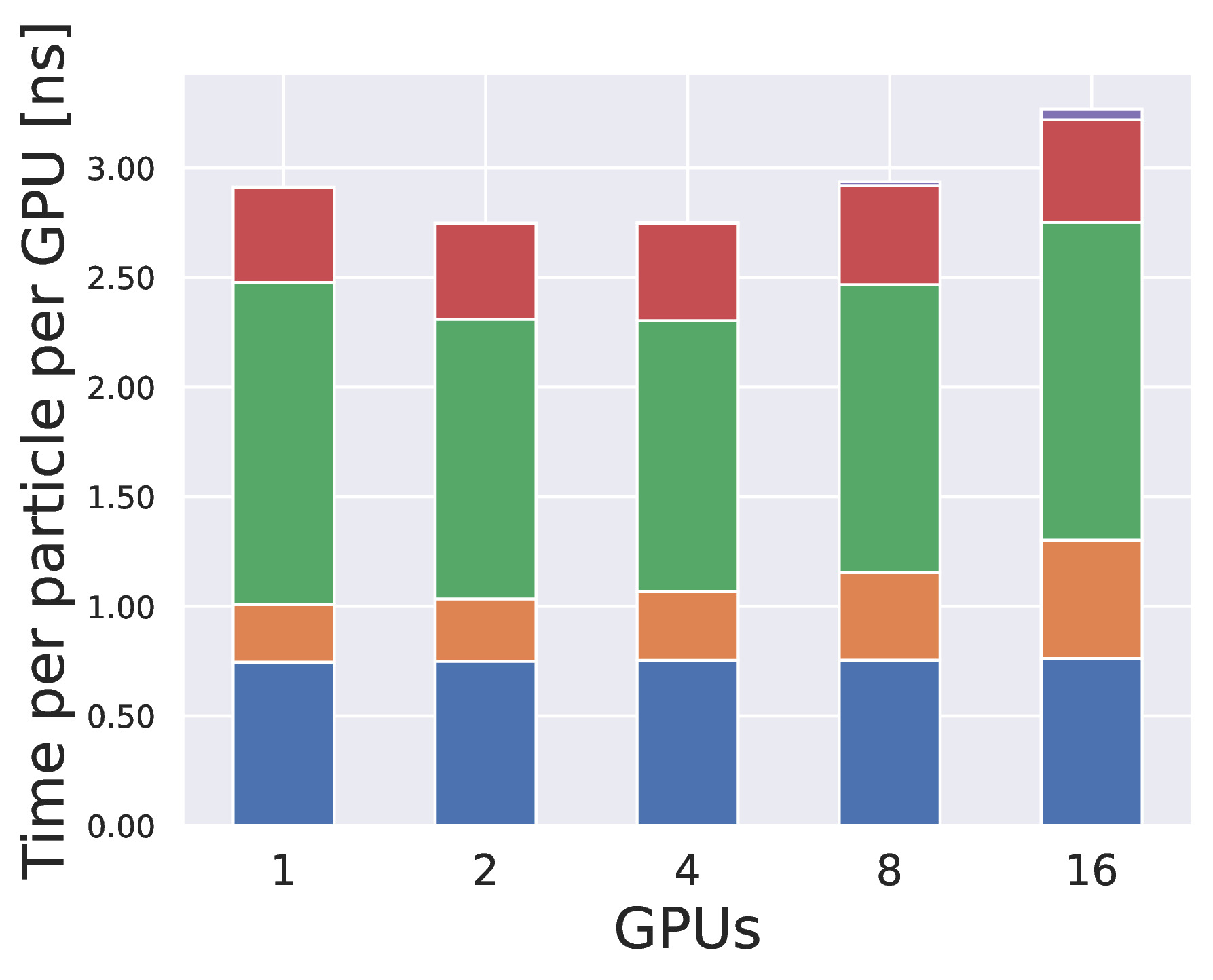}}%
  \subcaptionbox{Strong; $\rho = 100, N = 16e6$.\label{f:r100sa}}{\includegraphics[width=0.25\textwidth]{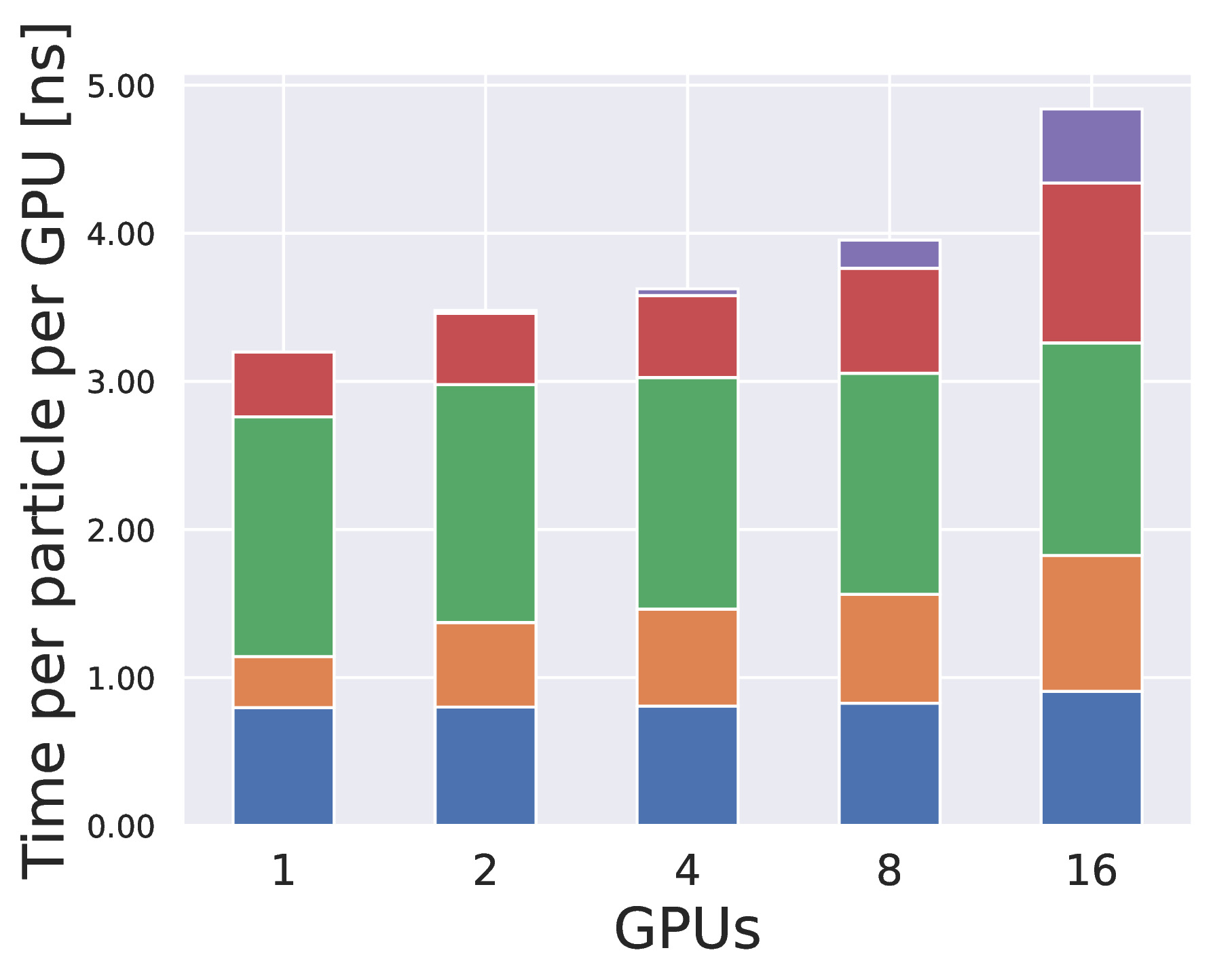}}%
  \subcaptionbox{Weak; $\rho = 100, n_P = 4e6$.\label{f:r100wa}}{\includegraphics[width=0.25\textwidth]{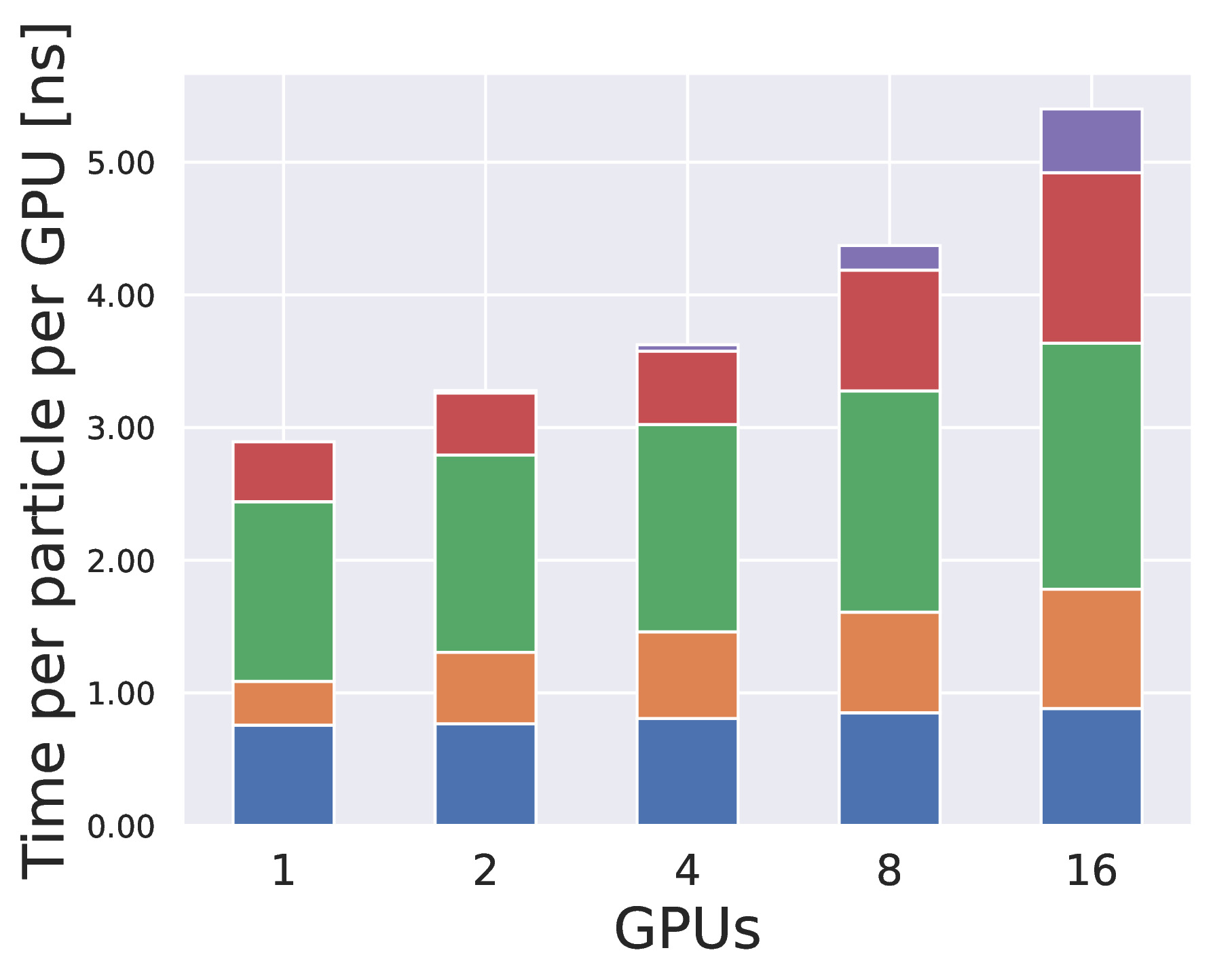}}%

  \caption{\revisedb{Normalized time in nanoseconds for strong and weak scaling runs using the PyKokkos kernels.  In particular, we report $\hat{T}_P = T_P/n_P = T_P P / N$, where  $T_P$ is the wall-clock time per simulation time step (averaged over 400 time steps);  $N$ is the total number of particles; $P$ is the number of GPUs which is equal to the number of MPI tasks,  and $n_P= N/P$ is the number of particles per GPU. Perfect scaling would be a horizontal line. It follows that the parallel efficiency $\eta_P =  \hat{T}_1/\hat{T}_P$, thus, the closer the plots are to a straight line the better the scaling. (In the weak scaling runs $n_P$ is fixed to $n$ for all $P$.) } Also, $\rho$ is the number of particles per cell. ``DSMC-Coll'' refers to all the kernels in Figure~\ref{fig_ek_atomic}; ``Recombination'' refers to C5 in Table~\ref{t:collisions}; and ``MPI'' refers to the overall MPI  communication costs. In the first row, we report timings for one DSMC step per PDE solve; in the second row, we report timings for 10 DSMC steps per PDE solve, as it is typically done in subcycled, multirate time-marching schemes in production  runs. For each run the total unnormalized time $T_P=\hat{T}_P n_P = \hat{T}_P N /P$. For example, in Figure~\ref{f:r1000s}, $T_8=6.5$ milliseconds and $T_{16}=4.1$ milliseconds.}
  \label{f:mpi}
\end{figure*}

\section{Overall scalability results}\label{s:results}

\MyPara{Setup}
All runs are in double precision and are performed using dedicated compute nodes. Each experiment is done with 5 simulations of 400 time steps. The timing we report is averaged across the runs, and we find the variance of each experiment to be less than 1\%. For all of the non-AMD runs, we used ``Lassen'',  a  system at Lawrence Livermore National Laboratory. Each node is equipped with an IBM Power9 CPU, 256GB of RAM, and four NVIDIA V100 Volta GPUs with 16 GB of RAM each. The V100 peak double precision performance is 7.8 TFLOPS and its memory bandwidth is 900 GB/sec. Lassen's infiniband interconnect's bandwidth is 12.5 GB/sec. For the strong and weak scaling experiments, we use one GPU per MPI task, with up to four MPI tasks per node. We use Python 3.8.8, CUDA 10.2, GCC 8.3.1, mpi4py 3.1.3, Numba 0.54.1, the ``develop'' branch of the PyKokkos GitHub repository, and the IBM Spectrum MPI.

For the runs to demonstrate the portability of PyKokkos, we used ``Tioga'', another system at Lawrence Livermore National Laboratory. Each node is equipped with an AMD Trento CPU, 512GB of RAM, and four AMD MI-250X GPUs with 32GB of RAM each. The MI-250X peak double precision performance is 45.0 TFLOPS and its memory bandwith is 3.2 TB/sec. For the experiments done on Tioga, we limit discussion to kernel performance on a single AMD GPU. We use Python 3.10, ROCm 5.4.3, HIP version 5.4.22804-474e8620, and the HIPCC compiler using AMD Clang 15.0.0 compiler.

\MyPara{Normalized time results}
In Figure~\ref{f:mpi} we report timings as a function of the number of GPUs for different problem sizes using Lassen. To present and discuss the results in Figure~\ref{f:mpi}, we introduce the following variables. Let $N$ be the number of particles (electrons). Let $P$ be the number of GPUs or equivalently MPI tasks as we use one GPU per MPI task. We define the grain size to be $n_P = N/P$ the number of particles per GPU. We report \emph{the normalized time} on $P$ GPUs defined by  $\hat{T}_P = T_P / n_P = T_P P / N$. Notice that for an embarrassingly parallel algorithm that has complexity $T_P=\bigO(N/P)$, we have $\hat{T}_P = \bigO(\hat{T}_1)$ so that all weak and strong scaling runs should be straight lines; however, our algorithm has overheads due to the MPI communication and the fact that the grid cells are replicated at each MPI task. We give details and a performance analysis further below.

In Figure~\ref{f:mpi} we report normalized times for the DSMC-Coll step, the recombination collision, the Coulombic step, the P2C step, and the MPI communication step.  All other costs are negligible.  The first row results were obtained with runs that use one DSMC-Coll, recombination, and Coulombic step per PDE solve.  The second row results were obtained with runs that  used a subcycled multirate scheme that takes ten DSMC-Coll / recombination / Coulombic steps per PDE solve~\citep{vahedi1995,bettencourt21}. Therefore for the second row runs, we have one P2C and {\tt MPI\_AllReduce()} every ten DSMC-Coll/recombination/Coulombic kernel calls. This explains the reduced costs for those two kernels. 

In Figure~\ref{f:r1000s},\ref{f:r1000w} we report strong and weak scaling results for $\rho=1000$; and Figure~\ref{f:r100s},\ref{f:r100w} show results for $\rho=100$. These timings depend on the number of cells (elements of finite-difference grid points). Similar definitions are used in Figure~\ref{f:r1000sa}-\ref{f:r100wa}.  Recall that in Figure~\ref{fig_ek_atomic}, we presented the breakdown for the stand-alone DSMC-Coll kernel (Table~\ref{t:pic}) without the recombination kernel. \emph{The total time in Figure~\ref{fig_ek_atomic} corresponds to the DSMC-Coll time in Figure~\ref{f:mpi}.}  For all strong scaling runs in Figure~\ref{f:r1000s},\ref{f:r100s}, we use $N=1.6e7$ particles; thus $M=160$K for $\rho=100$ and $M=1600$ for $\rho=1000$. For all  weak scaling runs we use $n=4e6$.

The parallel speedup on $P$ GPUs is defined as $s_P = T_1/T_P = \hat{T}_1 P / \hat{T}_P$. The parallel efficiency is defined as $\eta_P = s_P/P = \hat{T}_1/\hat{T}_P$. For example, in Figure~\ref{f:r1000sa}, $\hat{T}_1=2.53$ ns, $\hat{T}_8=2.99$ ns, and $\hat{T}_{16} = 3.70$ ns. Thus,  $\eta_8 \approx 85$\% and $\eta_{16} \approx 68$\%. In Figure~\ref{f:r100s},  $\hat{T}_1=3.21$ ns, $\hat{T}_8=5.83$ ns, and $\hat{T}_{16} = 9.41$ ns. Thus, $\eta_8 \approx 55$\% and $\eta_{16} \approx 34$\%.  What causes this efficiency drop especially in the small $\rho$ case? Next, we derive performance models of the four different kernels in Figure~\ref{f:mpi}, which we then use to interpret the observed efficiencies. As we will see, the main reason for the efficiency drop is the cell replication across MPI tasks.

\subsection{Performance Analysis}
\label{s:results:perf}

To better interpret the results, we first introduce performance models for steps reported in Figure~\ref{f:mpi}.
We define $T_{\mathsf{DSMC}}$, $T_{\mathsf{COUL}}$, $T_{\mathsf{MPI}}$, and $T_{\mathsf{P2C}}$, the wall-clock time for the DSMC-Coll kernel, the Coulombic kernel, the MPI communication, and the particle to cell kernel respectively. We define the normalize times $T_{\mathsf{DSMC}}=T_{\mathsf{DSMC}}/n_P$, $\hat{T}_{\mathsf{COUL}}=T_{\mathsf{COUL}}/n_P$, $\hat{T}_{\mathsf{MPI}}=T_{\mathsf{MPI}}/n_P$, and $\hat{T}_{\mathsf{P2C}} = T_{\mathsf{P2C}}/n_P$. The normalized times are the ones shown in Figure~\ref{f:mpi}. We omit discussion of the recombination kernel as the analysis is nearly identical to the P2C one.

\MyPara{The MPI communication kernel}
Since $\hat{T}_{\mathsf{MPI}}$ is the most prominent one, especially in the first row for $\rho=100$, we start with it. $T_{\mathsf{MPI}}$ is the cost of an all-reduce operation, that should scale as $\tau_c \log_2(P) M$, with $\tau_c$ being the inverse of internode bandwidth and assuming a hypercube topology with negligible latency costs~\citep{karypis03}. It follows that $\hat{T}_{\mathsf{MPI}} = \tau_c \log_2(P) P/\rho$. Therefore, the cost should (and does) decrease with increasing $\rho$ because the main work is done in the per-particle kernels, whereas the MPI communication scales with the grid size. Also the normalized per particle time increases with $P$ as the percentage of communication per particle increases. Thus, the formula for $\hat{T}_{\mathsf{MPI}}$ explains the scaling in the results that we observe.

We note that we do not present MPI scaling results on the AMD machine Tioga. The MPI implementation of our code would be identical, so differences observed in performance would be a result of differences in the underlying node communication structure or the system MPI software.

\begin{figure*}[t]
    \centering
    \includegraphics[width=1.0\textwidth]{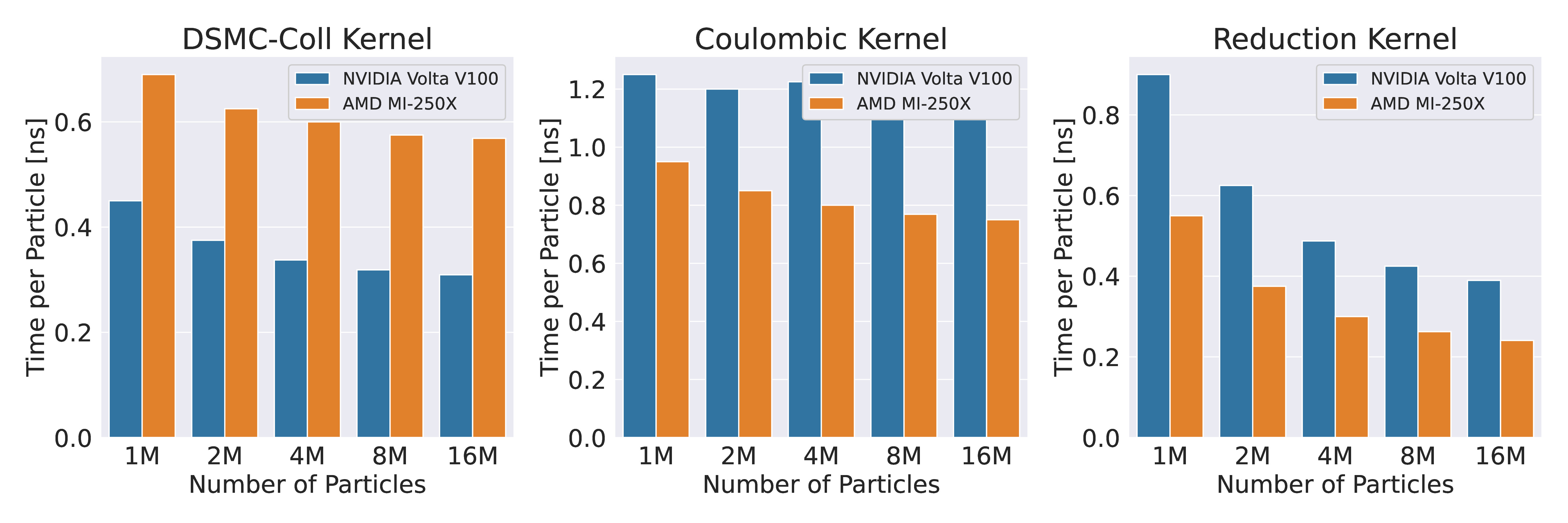}
    \centering
    \caption{\centering NVIDIA Volta V100 vs AMD MI-250X GPU runtime scalability results for three computational kernels.}
    \label{f:amdvsnv}
\end{figure*}

\MyPara{The DSMC-Coll kernel}
The DSMC-Coll kernel corresponds to the whole kernel analyzed in~\ref{fig_ek_atomic}. Here, we ignore the random number generation as it is an external library call. For the analysis we assume an infinite cache system in which the main cost is the loads and stores of the necessary data, followed by floating-point calculations. We ignore all costs related to atomic operations.  To present our analysis we define the following: $\tau_f$ is the \idef{machine time per double precision
  FLOP}, $\tau_m$ is the \idef{RAM access/byte}, $f_D=405$ is the
\idef{number of FLOPs/particle}, which includes advection as well as
collisions C1--C4 and $m_D=18\times8$  is the \idef{memory operations per
  particle}. This includes loads and stores for 3D positions, velocities, weights, and reads for random number and 3D electric
field, precomputed in the C2P step. The arithmetic intensity is
$f_D/m_D = 2.8$ and the machine imbalance is $\tau_m/\tau_f = 70$ for
double precision calculations on the NVIDIA V100. Then, the \idef{time
  per particle} is
$T_{\mathsf{DSMC}} = f_D \tau_f + m_D \tau_D $, $  \approx  \tau_m m_D$.
For $\tau_m = 1/900 $ ns, the estimated cost per particle is 0.16 ns. In~\ref{fig_ek_atomic}, we observe roughly 0.42 ns. We attribute the discrepancy to register pressure; the compiler shows about 96 registers per thread, which leads to reduced occupancy. Restricting the number of registers helps, but the benefits are not as dramatic due to register spilling. There is also some sensitivity on the number of blocks and threads per block.


\MyPara{P2C kernel}
  Recall that in the P2C kernel we need to do four segmented reductions on particle properties, where the number of segments is equal to the number of field cells.  In our implementation, the P2C kernel requires $\bigO(1)$ work per particle to find its cell and then $O(M)$ concurrent reductions of approximately $M/n_P$ size each. The computation for the first part is negligible and we ignore it. The main computation cost is managing the $O(M)$ atomics-based reductions. 
  
The memory bandwidth cost $m$ is loading the per-particle properties and thus,  $m=4\times 8 $ \idef{double precision memory operations  per particle}. In all, we obtain $\hat{T}_{\mathsf{P2C}} = m \tau_m + O(M/n_P) = m \tau_m + O(P/\rho)$. 

The $m \tau_m$ cost is $0.03$ ns  and thus negligible compared to the observed $\hat{T}_{\mathsf{P2C}}$. For example, in Figure~\ref{f:r1000s}, we observe $\hat{T}_{\mathsf{P2C}}=0.4$ ns for $P=1$. Thus the dominant term is the atomic reductions cost $O(P/\rho)$. This analysis explains the slight increase of $\hat{T}_{\mathsf{P2C}}$ with increasing $P$, which is amplified with smaller $\rho$.

\MyPara{Coulombic kernel} The Coulombic kernel can be split into three components - random number generation, pair assignment, and collision execution. As this is our most expensive kernel, we analyze and measure the individual times for each of these three steps to further understand the breakdown of time spent. As this is an external library call, we ignore it here but note that the RNG cost is no more than 10\% of the computation time of this kernel.

The second step is the creation of the lists $P_j$ of particles per grid cell. Many other codes store particles in a sorted manner. With this structure, it is substantially simpler to set up electron-electron collisions between particles of the same cell. However, this method poses its own challenges in regards to data management and communication between processes.

Our creation of $\{P_j\}$ requires a simple loop over all the particles $l$ to determine their cell and write the index of particle $l$ into our pairing array $P$. Because this kernel is a simple $O(N)$ loop with just one memory read and write per particle, the atomic operations takes up the bulk of the computation time, which can be up to $40$\% of the time of this kernel.

The final step, which comprises the bulk of the Coulombic kernel runtime, performs the collision compuatations. The DSMC-Coll kernel is able to use coalesced memory reads and writes due to its embarrassingly parallel nature (aside from particle creation, a somewhat rare process). At each timestep, the Coulombic kernel reads indices from adjacent memory locations in $P$, however these indices refer to ``random'' locations within our data structures. We then perform $\frac{N}{2}$ collisions between particles randomly stored within the unsorted data lists. For each particle pair, 6 reads and 6 writes are performed. The non-coalesced memory reads and writes are the primary cause of this relatively high runtime of this kernel piece compared to the DSMC-Coll kernel.

\MyPara{Summary} Within $O(1)$ factors we can explain the scaling and absolute numbers of our code. The degradation in scalability is due to the field mesh replication. However, our solution is simple and for many practical problems of interest the multi-GPU version offers significant speedups especially for large $\rho$ and subcycled time marching schemes.

\subsection{Extending to AMD GPUs} \label{s:results:amd}

We now discuss the results of runs done on both NVIDIA Volta V100 and AMD MI-250X GPUs. This is done as a demonstration of the portability of PyKokkos and a comparison between the two specific GPUs, rather than a comprehensive comparison of the two architectures. Recall that PyKokkos allows us to use the exact same code base for both machines. We run our Boltzmann solver on both devices for a number of different problem sizes. The results can be seen in Figure~\ref{f:amdvsnv}. We consider problem sizes from 1 to 16 million particles; for all cases we use $16,000$ grid cells.

Recall from Section~\ref{s:results} that the MI-250X reports over 5$\times$ better peak double precision performance and over 3$\times$ better memory bandwith. However, these results are drawn from highly optimized test problems; applications rarely see such performance. Still, we would thus expect to see better runtimes on our MI-250X. This holds true for the Coulombic and reduction kernels, however we observe the DSMC-Coll kernel performs better on the Volta V100.

We know from our previous analysis in Section~\ref{s:results:perf} that the DSMC-Coll kernel is memory bound, not compute bound. We thus focus our analysis and testing on memory operations: atomics; register pressure; reads/writes; cache effects. We probe specifically these aspects using two modified versions of the DSMC-Coll kernel. The first modification we consider is the removal of the atomic operation used to track write indices for no particles by setting the probability of ionization collisions to $0$. The second test involves reducing the number of registers used within the kernel. As discussed in Section~\ref{s:results:perf}, the compiler shows about 96 registers per thread. A significant number of these registers are used in the cross section calculations; we can use artificial analytic cross sections to eliminate the use of these registers.

\begin{figure}[H]
    \centering
    \includegraphics[width=0.5\textwidth]{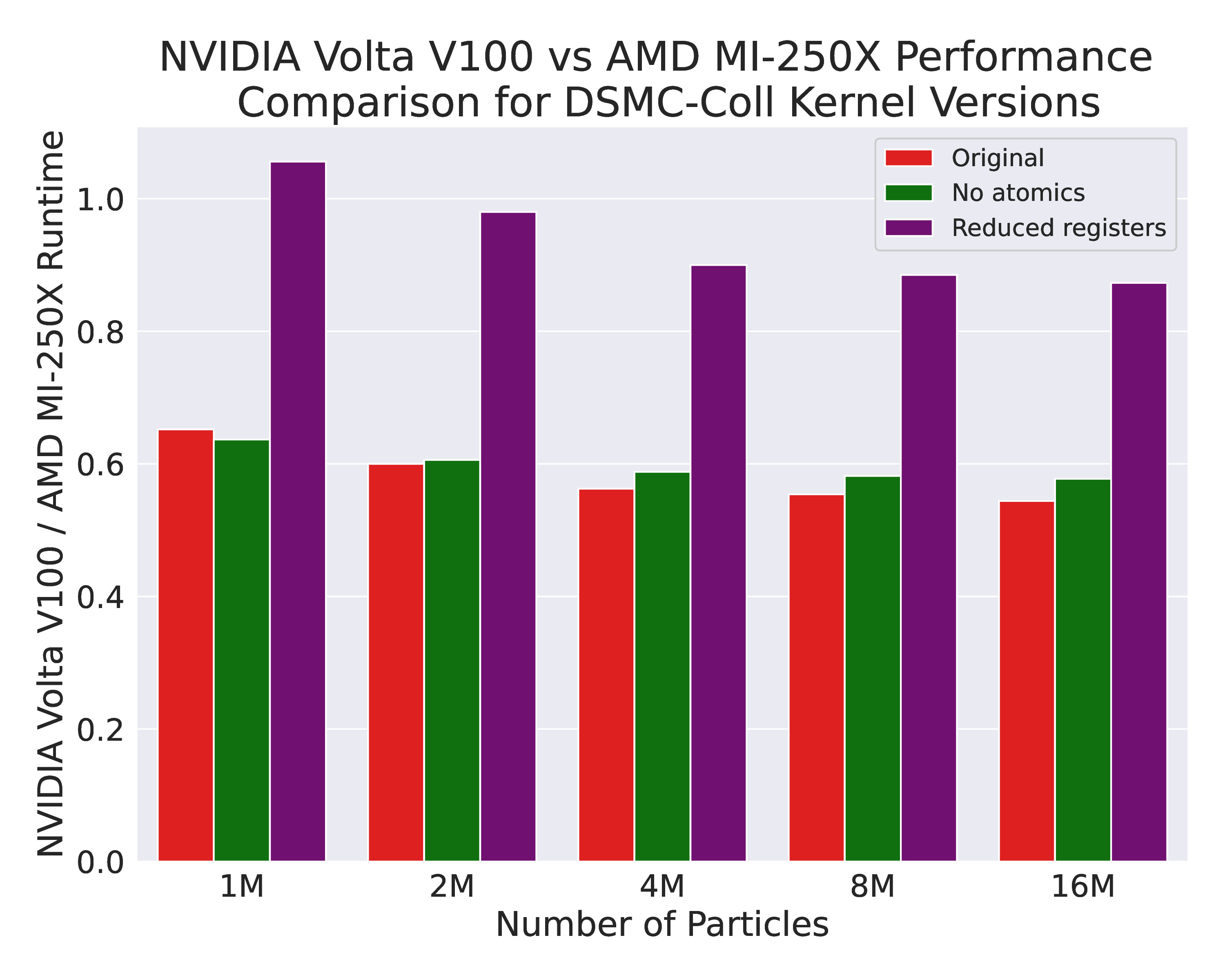}
    \caption{Relative performance on NVIDIA Volta V100 vs. AMD MI-250X for the original and two modified versions of the DSMC-Coll kernel. Ratios less than one correspond to better performance on the Volta V100.}
    \label{f:nvamdadjust}
\end{figure}

The results of these experiments can be seen in Figure~\ref{f:nvamdadjust}; we used the same problem sizes as in the previous comparisons. Here we report the ratio of the runtime on the NVIDIA Volta V100 to the runtime on the AMD MI-250X. Values less than 1 correspond to better performance on the V100; values greater than 1 correspond to better performance on the MI-250X. The relative performance on the original and the version with no atomics is nearly the same; we conclude that the use of atomics is not the cause for the performance discrepancy. We do, however, see a significant difference in relative performance using the version of the kernel with no reduced registers. We thus find that the MI-250X is significantly more sensitive to high register counts, which contributes to its poor relative performance.  While this experiment reduces the number of registers used, even this version of the DSMC-Coll kernel uses substantially more than the Coulombic or reduction kernels; this is consistent with our findings. As an aside, we note that compiler options for both NVIDIA and AMD GPUs can allow the user to specify the number of registers per thread. This is beyond the scope of this paper; we chose to use our software as given with no internal modifications.

While the reduced register usage example improved the relative performance of the DSMC-Coll kernel on the AMD MI-250X, the results are still far from what we see for the Coulombic and reduction kernels. We believe the performance shown for the DSMC-Coll can also be attributed to the difference in speed of memory operations on the two GPUs. We consider a simple test using an artificial kernel to further understand this behavior. 

Figure~\ref{fig:proxy} shows a simple kernel meant to serve as a proxy to our other kernels. It contains a for loop that runs over all indices for a certain number of iterations. Line 4 reads in the values in $y$; line 5 then sets the value of $x$ using a new value. Line 6 sets the value of a register $t$; here, values in $y$ can be grabbed from the cache. Lines 7-8 grab the value of $x$ previously set, then overwrite it using the register $t$. This kernel mimics the steps of the electron kernel -- particles are first read in, data is assigned to registers, and arrays are written to either as overwrites or in new locations. By varying the sizes of the arrays and the number of iterations in the for loop, we can compare different aspects of the performance across NVIDIA and AMD GPUs.
1z
\begin{figure}[H]
    \centering
      \begin{small}
      \begin{algorithmic}[1]
      \Function{proxyKernel}{$x, y, size, iterations$}
        \State $tid \leftarrow$ \Call{$getTID$}{}
    
        \For{$i$ in \Call{$range$}{$iterations$}}
          \State $x[i] \leftarrow y[i]$
          \State $x[i] \leftarrow 1 + i / 10$
          \State $t \leftarrow y[i]$
          \State $y[i] \leftarrow x[i]$
          \State $x[i] \leftarrow t$
        \EndFor
        \EndFunction
      \end{algorithmic}
      \end{small}
    \caption{Proxy Kernel to Compare AMD MI-250X and NVIDIA Volta V100~\label{fig:proxy}}
\end{figure}

\begin{table}[h]
\centering
        \begin{tabular}{|l|l|l|l|l|l|}
        \cline{1-6}
        \multirow{2}{*}{Iterations} & \multicolumn{5}{c|}{Message Size} \\
        \cline{2-6}
         & \textbf{1KB} & \textbf{10KB} & \textbf{100KB} & \textbf{1MB} & \textbf{10MB}\\
        \hline
        \textbf{1} & 2.13 & 1.86 & 2.05 & 4.92 & 5.09\\
        \cline{1-6}
        \textbf{10} & 0.89 & 5.80 & 8.69 & 8.63 & 8.70\\
        \cline{1-6}
        \textbf{100} & 0.97 & 6.83 & 10.26 & 12.63 & 13.21\\
        \cline{1-6}
        \textbf{1000} & 0.91 & 2.91 & 8.73 & 5.74 & 4.46\\
        \cline{1-6}
        \textbf{10000} & 0.43 & 0.79 & 2.50 & 1.61 & 1.26\\
        \cline{1-6}
        \end{tabular}
    \caption{Relative performance of proxy kernel, shown as runtime on NVIDIA Volta V100 / runtime on AMD MI-250X, as a function of message size and performance. We observe that the Volta V100 performs better for a high number of iterations and smaller message sizes, whereas the MI-250X performs better for larger message sizes and fewer iterations. ~\label{fig:proxy:results}}
\end{table}

Table~\ref{fig:proxy:results} shows the results of our experimentation with
the proxy kernel. The first column shows the number of iterations and
every other column shows the ratio of the total time taken when
running the kernel on the NVIDIA GPU over the total time taken when
running on the AMD GPU. In other words, this shows the speedup we
obtain using the AMD GPU. The results show that the performance
difference between the two GPUs varies greatly depending on both the
size and the number of iterations and is highly unpredictable. For
100 iterations, the AMD GPU is over 6$\times$ faster for arrays larger
than 1KB, reaching up to a 13$\times$ speedup for our largest array
size. Meanwhile, the NVIDIA GPU performed better for our smallest
array size with the number of iterations being greater than 1. Further investigation here is ongoing research. 

\section{Simulation of a Glow Discharge Apparatus}

We use our solver to simulate a low-pressure  RF glow discharge apparatus, in which an AC voltage is applied across a discharge tube. For the regime of glow discharge problems we study here, ionization fractions are no larger than $10^{-6}$. At such low ionization fractions, Coulombic collisions have almost no effect on the relevant physics. As a result, glow discharge simulations typically omit the modeling of Coulombic collisions. Our implementation is done in part to investigate the HPC aspect of Coulombic collisions. It also allows for future extensions to different physical problems in which these collisions become an integral piece of physics.

\begin{figure}[ht]
    \centering
    \includegraphics[scale=0.35]{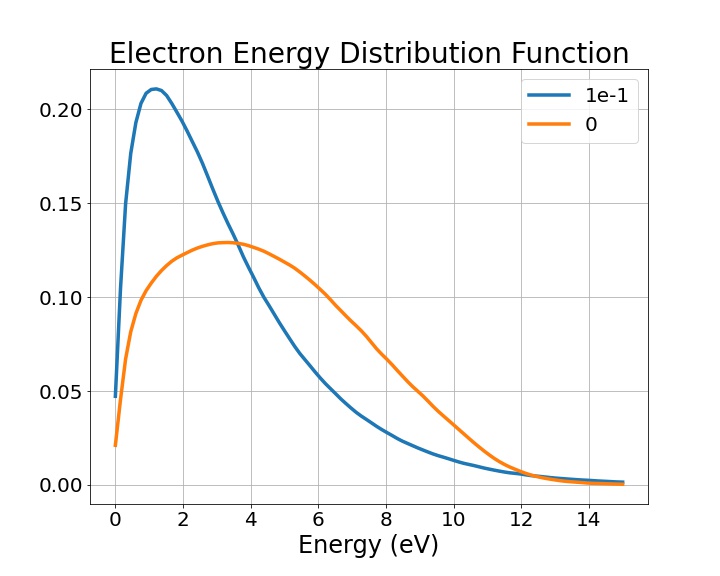}
    \caption{Electron energy distribution functions of 0-D steady-state simulations for electric field $E = 32.2 \mathrm{V} / \mathrm{cm}$, ionization fractions $n_e / n_n = 0$ and $n_e / n_n = 0.1$.}
    \label{physicsresults_image}
\end{figure}

To explore this further, we consider a 0-D test problem in which we only consider the DSMC part of the solver. We consider a constant electric field and fixed densities for all species, including electrons. We then step forward in time until the observable parameters (electron temperature, reaction rates, and mobility) become constant. This simulation to steady state is a standard problem in kinetic theory \citep{Hagelaar_2005}. With this problem, we can easily alter the ionization fraction to demonstrate the effects of modeling Coulombic collisions. In Figure~\ref{physicsresults_image} we show the steady-state electron energy distribution functions for two simulations. These simulations used a neutral density $n_n = 3.22 \times 10^{16} \frac{\#}{\mathrm{cm}^3}$ and electric field $E = 32.2 \frac{\mathrm{V}}{\mathrm{cm}}$. For simplicity, we consider a case with just two collisions -- elastic and ionization -- along with the Coulombic interactions. Shown are the results for cases with ionization fractions of $\frac{n_e}{n_n} = 0$ and $\frac{n_e}{n_n} = 10^{-1}$. The Coulombic collisions play a clear role in altering the behavior of the system. For example, the electron temperatures for the two cases are approximately $3.43\mathrm{eV}$ and $4.72\mathrm{eV}$; the reaction rates and calculated electron mobility differ on a similar scale. 

\color{black}

\section{Related Work}\label{s:related}

Comparison of our code to other LTP codes and discussion on related work is discussed extensively in previous work~\citep{almgrenbell-e22}. We compared to other particle codes without including Coulombic collisions. We reported double precision performance of 1.4 ns/particle/time step, which amounts to over 5x improvement of what we believe to be the state of the art. With 16 GPUs, we achieve a total of 0.2 ns/particle/time step. Here, we focus on related work on Coulombic interactions. There is relatively minimal work on GPU algorithms for Coulombic collisions in the context of the problem we discuss. Van Gorp et al. present their code Simbuca that utilizies GPUs to simulate Coulombic interactions between ions \citep{Simbuca}. They solve an approximate N-body problem, which is unnecessary for the simplified model we consider. Zhao models Coulombic collisions in an ionized plasma using Nanbu's method, a common alternative to the method of Takizuka and Abe that we employ. Their performance discussion is limited to implementing the Nanbu method on one CPU \citep{zhao2018}. Other DSMC codes such as SPARTA model collisions with particles that are pre-sorted spatially. The trivial extension to Coulombic collision is clear \citep{sparta}, however the computational cost of these particle-particle collisions is greatly reduced due to the simplified partner selection step. This of course comes with the additional cost of regular sorting of particles, which includes synchronizations, data transfers, and more. As such, runtimes are not directly comparable between these two codes. Other work on Coulombic collisions for an electron Boltzmann solver focuses on deterministic codes that solve for the electron distribution function $f$ directly \citep{Hagelaar_2016, jeongyoung}.

 
\section{Conclusion}
To our knowledge this is the first multi-GPU solver for LTPs and the first to report details for recombination and Coulombic collisions. We presented a simple but effective MPI algorithm for up to 64M particles and 160K cells. Using Python and PyKokkos, we were able to rapidly perform algorithmic exploration and implementation. We employ the performance portability of PyKokkos to report and analyze performance results on both NVIDIA and AMD GPUs. We also dive into understanding the performance discrepancies between the two GPUs. Future work includes using particle reweighing to adaptively control the approximation of the velocity distribution function; extending the MPI implementation to partitioned $\Omega$; and using multi-resolution methods to accelerate convergence to a steady state. One of the main limitations of our work is that we only consider uniformly weighted particles. In certain plasma regimes, such uniform weighting may require a large number of samples. We are currently working on extending the described algorithms to variable weight particle methods.

\begin{funding}
    This material is based upon work supported by the Department of Energy, National Nuclear Security Administration under Award Number DE-NA0003969.
\end{funding}

\bibliographystyle{SageH}
\bibliography{main.bib}

\begin{thebibliography}{28}
\providecommand{\natexlab}[1]{#1}
\providecommand{\url}[1]{\texttt{#1}}
\providecommand{\urlprefix}{URL }
\expandafter\ifx\csname urlstyle\endcsname\relax
  \providecommand{\doi}[1]{DOI:\discretionary{}{}{}#1}\else
  \providecommand{\doi}{DOI:\discretionary{}{}{}\begingroup \urlstyle{rm}\Url}\fi

\bibitem[{Al~Awar et~al.(2021)}]{pykokkos21}
Al~Awar N et~al. (2021) A performance portability framework for {P}ython.
\newblock In: \emph{International Conference on Supercomputing}.

\bibitem[{Almgren-Bell et~al.(2022)Almgren-Bell, Awar, Geethakrishnan, Gligoric and Biros}]{almgrenbell-e22}
Almgren-Bell J, Awar NA, Geethakrishnan DS, Gligoric M and Biros G (2022) A multi-gpu python solver for low-temperature non-equilibrium plasmas.
\newblock In: \emph{2022 IEEE 34th International Symposium on Computer Architecture and High Performance Computing (SBAC-PAD)}. pp. 140--149.
\newblock \doi{10.1109/SBAC-PAD55451.2022.00025}.

\bibitem[{Alves et~al.(2018)}]{alves-turner18}
Alves L et~al. (2018) Foundations of modelling of nonequilibrium low-temperature plasmas.
\newblock \emph{Plasma Sources Science and Technology} 27(2).

\bibitem[{Bauer et~al.(2019)Bauer,  et~al.}]{bauer2019code}
Bauer M,  et~al. (2019) Code generation for massively parallel phase-field simulations.
\newblock In: \emph{International Conference for High Performance Computing, Networking, Storage and Analysis}.

\bibitem[{Bettencourt et~al.(2021)}]{bettencourt21}
Bettencourt M et~al. (2021) {EMPIRE-PIC}: a performance portable unstructured particle-in-cell code.
\newblock \emph{Communications in Computational Physics} 30(SAND-2021-2806J).

\bibitem[{Blelloch(1996)}]{blelloch1996}
Blelloch GE (1996) Programming parallel algorithms.
\newblock \emph{Communications of the ACM} 39(3).

\bibitem[{Grama et~al.(2003)Grama, Gupta, Karypis and Kumar}]{karypis03}
Grama A, Gupta A, Karypis G and Kumar V (2003) \emph{An Introduction to Parallel Computing: Design and Analysis of Algorithms}.
\newblock Second edition. Addison Wesley.

\bibitem[{Hagelaar(2015)}]{Hagelaar_2016}
Hagelaar GJM (2015) Coulomb collisions in the boltzmann equation for electrons in low-temperature gas discharge plasmas.
\newblock \emph{Plasma Sources Science and Technology} 25(1): 015015.
\newblock \doi{10.1088/0963-0252/25/1/015015}.
\newblock \urlprefix\url{https://dx.doi.org/10.1088/0963-0252/25/1/015015}.

\bibitem[{Hagelaar and Pitchford(2005)}]{Hagelaar_2005}
Hagelaar GJM and Pitchford LC (2005) Solving the boltzmann equation to obtain electron transport coefficients and rate coefficients for fluid models.
\newblock \emph{Plasma Sources Science and Technology} 14(4): 722.
\newblock \doi{10.1088/0963-0252/14/4/011}.
\newblock \urlprefix\url{https://dx.doi.org/10.1088/0963-0252/14/4/011}.

\bibitem[{Harris et~al.(2020)}]{HarrisETAL20Numpy}
Harris CR et~al. (2020) Array programming with {NumPy}.
\newblock \emph{Nature} 585(7825).

\bibitem[{Hur et~al.(2019)}]{hurplasma_gpu19}
Hur MY et~al. (2019) Model description of a two-dimensional electrostatic particle-in-cell simulation parallelized with a graphics processing unit for plasma discharges.
\newblock \emph{Plasma Research Express} 1(1).

\bibitem[{Ji et~al.(2021)Ji, Lee, Held and Yun}]{jeongyoung}
Ji JY, Lee MU, Held ED and Yun GS (2021) {Moments of the Boltzmann collision operator for Coulomb interactions}.
\newblock \emph{Physics of Plasmas} 28(7): 072113.
\newblock \doi{10.1063/5.0054457}.
\newblock \urlprefix\url{https://doi.org/10.1063/5.0054457}.

\bibitem[{Kongpiboolkid and Mongkolnavin(2015)}]{kongpiboolkid2015}
Kongpiboolkid W and Mongkolnavin R (2015) {Plasma characteristics of argon glow discharge produced by AC power supply operating at low frequencies}.
\newblock \emph{AIP Conference Proceedings} 1657(1): 150004.
\newblock \doi{10.1063/1.4915243}.
\newblock \urlprefix\url{https://doi.org/10.1063/1.4915243}.

\bibitem[{Lam et~al.(2015)}]{lam2015numba}
Lam SK et~al. (2015) Numba: A {LLVM}-based {P}ython {JIT} compiler.
\newblock In: \emph{Second Workshop on the LLVM Compiler Infrastructure in HPC}.

\bibitem[{lxcat()}]{lxcat}
lxcat (2021) {Phelps} and {TRINITI} databases.
\newblock \url{www.lxcat.net}.

\bibitem[{Okuta et~al.(2017)}]{nishino2017cupy}
Okuta R et~al. (2017) {CuPy}: A {NumPy}-compatible library for {NVIDIA GPU} calculations.
\newblock In: \emph{Workshop on Machine Learning Systems (LearningSys) in The Thirty-first Annual Conference on Neural Information Processing Systems (NIPS)}.

\bibitem[{Oliphant(2007)}]{Oliphant07ScientificPython}
Oliphant TE (2007) Python for scientific computing.
\newblock \emph{Computing in Science and Engineering} 9(3).

\bibitem[{Plimpton et~al.(2019)Plimpton, Moore, Borner, Stagg, Koehler, Torczynski and Gallis}]{sparta}
Plimpton SJ, Moore SG, Borner A, Stagg AK, Koehler TP, Torczynski JR and Gallis MA (2019) {Direct simulation Monte Carlo on petaflop supercomputers and beyond}.
\newblock \emph{Physics of Fluids} 31(8).
\newblock \doi{10.1063/1.5108534}.
\newblock \urlprefix\url{https://doi.org/10.1063/1.5108534}.
\newblock 086101.

\bibitem[{Takizuka and Abe(1977)}]{TAoriginal}
Takizuka T and Abe H (1977) A binary collision model for plasma simulation with a particle code.
\newblock \emph{Journal of Computational Physics} 25(3): 205--219.
\newblock \doi{https://doi.org/10.1016/0021-9991(77)90099-7}.
\newblock \urlprefix\url{https://www.sciencedirect.com/science/article/pii/0021999177900997}.

\bibitem[{Trott et~al.(2021)}]{TrottETAL21Kokkos}
Trott C et~al. (2021) The {K}okkos ecosystem: Comprehensive performance portability for high performance computing.
\newblock \emph{Computing in Science and Engineering} 23(5).

\bibitem[{Vahedi and Surendra(1995)}]{vahedi1995}
Vahedi V and Surendra M (1995) A {M}onte {C}arlo collision model for the particle-in-cell method: applications to argon and oxygen discharges.
\newblock \emph{Computer Physics Communications} 87(1-2).

\bibitem[{{Van Gorp} et~al.(2011){Van Gorp}, Beck, Breitenfeldt, {De Leebeeck}, Friedag, Herlert, Iitaka, Mader, Kozlov, Roccia, Soti, Tandecki, Traykov, Wauters, Weinheimer, Zákoucký and Severijns}]{Simbuca}
{Van Gorp} S, Beck M, Breitenfeldt M, {De Leebeeck} V, Friedag P, Herlert A, Iitaka T, Mader J, Kozlov V, Roccia S, Soti G, Tandecki M, Traykov E, Wauters F, Weinheimer C, Zákoucký D and Severijns N (2011) Simbuca, using a graphics card to simulate coulomb interactions in a penning trap.
\newblock \emph{Nuclear Instruments and Methods in Physics Research Section A: Accelerators, Spectrometers, Detectors and Associated Equipment} 638(1): 192--200.
\newblock \doi{https://doi.org/10.1016/j.nima.2010.11.032}.
\newblock \urlprefix\url{https://www.sciencedirect.com/science/article/pii/S0168900210024940}.

\bibitem[{Villani(2002)}]{villani2002review}
Villani C (2002) A review of mathematical topics in collisional kinetic theory.
\newblock \emph{Handbook of mathematical fluid dynamics} .

\bibitem[{Virtanen et~al.(2020)}]{VirtanenETAL20Scipy}
Virtanen P et~al. (2020) {{SciPy} 1.0: Fundamental Algorithms for Scientific Computing in {P}ython}.
\newblock \emph{Nature Methods} 17.

\bibitem[{Wang et~al.(2008)Wang, Lin, Calfish, Cohen and Dimits}]{TAcomparison}
Wang C, Lin T, Calfish R, Cohen B and Dimits A (2008) Particle simulation of coulomb collisions: Comparing the methods of takizuka \& abe and nanbu.
\newblock \emph{J. Comp. Phys.} 227: 4308--.
\newblock \doi{10.1016/j.jcp.2007.12.027}.

\bibitem[{Zhang et~al.(2021)Zhang, Myers, Gott, Almgren and Bell}]{amrex21}
Zhang W, Myers A, Gott K, Almgren A and Bell J (2021) {AMR}e{X}: Block-structured adaptive mesh refinement for multiphysics applications.
\newblock \emph{The International Journal of High Performance Computing Applications} 35(6).

\bibitem[{Zhao(2018)}]{zhao2018}
Zhao Y (2018) {A binary collision Monte Carlo model for temperature relaxation in multicomponent plasmas}.
\newblock \emph{AIP Advances} 8(7): 075016.
\newblock \doi{10.1063/1.5035362}.
\newblock \urlprefix\url{https://doi.org/10.1063/1.5035362}.

\bibitem[{Ziogas et~al.(2021)}]{ziogas2021productivity}
Ziogas AN et~al. (2021) Productivity, portability, performance: Data-centric {P}ython.
\newblock In: \emph{International Conference for High Performance Computing, Networking, Storage and Analysis}.

\end{thebibliography}

\end{document}